\documentclass[12pt]{article}
\pdfoutput=1

\usepackage{amsmath,amssymb,multirow}
\usepackage{hyperref}
\usepackage{fullpage}

\allowdisplaybreaks[1]

\newtheorem{theorem}{Theorem}

\begin{document}

\title{A brute-force search for\\
       R-symmetric Wess-Zumino models}
\author{James Brister\textsuperscript{a,\dag},
        Shihao Kou\textsuperscript{a,\ddag},
        Zhengyi Li\textsuperscript{a,\S},
        Longjie Ran\textsuperscript{a,\P}
        Zheng Sun\textsuperscript{a,b,*}\\
        \normalsize\textsuperscript{a}%
                   \textit{College of Physics, Sichuan University,}\\
        \normalsize\textit{29 Wangjiang Road, Chengdu 610064, P.\ R.\ China}\\
        \normalsize\textsuperscript{b}%
                   \textit{CAS Key Laboratory of Theoretical Physics,}\\
        \normalsize\textit{Institute of Theoretical Physics, Chinese Academy of Sciences,}\\
        \normalsize\textit{55 Zhongguancun East Street, Beijing 100190, P.\ R.\ China}\\
        \normalsize\textit{E-mail:}
                   \parbox[t]{31em}{
                   \textsuperscript{\dag}\texttt{jbrister@scu.edu.cn,}
                   \textsuperscript{\ddag}\texttt{1786957287@qq.com,}
                   \textsuperscript{\S}\texttt{897237485@qq.com,}
                   \textsuperscript{\P}\texttt{2019222020004@stu.scu.edu.cn,}
                   \textsuperscript{*}\texttt{sun\_ctp@scu.edu.cn}}
       }
\date{}
\maketitle

\begin{abstract}

This work makes an exhaustive search for generic renormalizable R-symmetric Wess-Zumino models with up to $5$ chiral fields, and checks the consistency of their vacuum solutions with predictions from the Nelson-Seiberg theorem and its generalizations.  Each model is recorded as the R-charge assignment of fields, which uniquely determines the cubic polynomial superpotentials with generic coefficients.  Redundancy from permutation symmetries and reducible models are properly eliminated in the searching algorithm.  We found that among $859$ models in total, $19$ of them have supersymmetric vacua unpredicted by the Nelson-Seiberg theorem and its generalizations.  These exceptional models have their specific R-charge assignments covered by constructions found in previous literature.  The search result can be used to estimate the accuracy of the field counting method for finding supersymmetric models in the string landscape.  More applications of the dataset are expected in future work.

\end{abstract}

\section{Introduction}

The Nelson-Seiberg theorem~\cite{Nelson:1993nf} and its extensions~\cite{Kang:2012fn, Li:2020wdk} play important roles in 4-dimensional $\mathcal{N} = 1$ supersymmetry (SUSY) models~\cite{Nilles:1983ge, Martin:1997ns, Baer:2006rs, Terning:2006bq, Intriligator:2007cp}, which are extensively used in the study of new physics beyond the Standard Model.  According to the revised Nelson-Seiberg theorem~\cite{Kang:2012fn}, SUSY breaking happens in a Wess-Zumino models with a generic polynomial superpotential if and only if the superpotential has an R-symmetry and there are more R-charge $2$ fields than R-charge $0$ fields.  In addition, metastable SUSY breaking vacua can be obtained from models with approximate R-symmetries~\cite{Intriligator:2006dd, Intriligator:2007py}.  Although SUSY runaway directions related to R-symmetries or non-R $U(1)$ symmetries are common in SUSY breaking models~\cite{Ferretti:2007ec, Ferretti:2007rq, Azeyanagi:2012pc, Sun:2018hnk}, phenomenologically plausible models can still be built to avoid runaway to infinite field values in a cosmological time scale.  Finally, after coupling to the mediation sector, R-symmetries need to be broken to obtain a proper mass spectrum for SUSY particles \cite{Shih:2007av, Carpenter:2008wi, Sun:2008va, Komargodski:2009jf, Curtin:2012yu, Liu:2014ida, Fu:2023sfk}.  Hence constructing SUSY breaking models with R-symmetries takes the first step towards SUSY phenomenology models.

On the other hand, if the superpotential has an R-symmetry, and the number of R-charge 2 fields is less than or equal to the number of R-charge 0 fields, generic Wess-Zumino models give SUSY vacua~\cite{Sun:2011fq} with vanishing superpotentials~\cite{Kappl:2008ie, Dine:2009sw, Brister:2021xca}, which also become SUSY vacua in supergravity (SUGRA) extension of these models.  This part of the theorem is also true for discrete and non-Abelian R-symmetries~\cite{Brister:2022xxx}, and has applications in flux compactifications of type IIB superstring theory~\cite{Grana:2005jc, Douglas:2006es, Blumenhagen:2006ci, Ibanez:2012zz}.  By identifying R-symmetries with geometrical symmetries of the Calabi-Yau manifold and turning on R-invariant fluxes, a large number of SUSY vacua with zero cosmological constants can be produced at the string scale~\cite{DeWolfe:2004ns, DeWolfe:2005gy, Dine:2005gz, Palti:2007pm, Kanno:2017nub, Palti:2020qlc, Kanno:2020kxr}.  SUSY breaking and the cosmological constant are then non-perturbatively generated at lower scales, notably through the racetrack mechanism~\cite{Krasnikov:1987jj, Casas:1990qi, Taylor:1990wr, deCarlos:1992kox, Dine:1999dx}.  Such SUSY vacua compose the third branch of the string landscape which prefers low-scale SUSY breaking~\cite{Dine:2004is, Dine:2005yq}. And the low energy effective Wess-Zumino models with R-symmetries provide a simple way to study flux compactification models on this branch.

In both of the previous applications, whether a model can have a SUSY vacuum or not can be determined from counting and comparing R-charge $2$ and R-charge $0$ fields.  Assuming genericity, no explicit calculation for the actual vacua is needed, thus it is possible to conduct a fast survey of vacua in the landscape.  But the genericity assumption may affect the accuracy of the field counting method.  For a specific model, genericity in coefficients means that the property of vacua withstands small random perturbations on coefficients.  This type of non-generic models only composes a null set in the coefficient parameter space.  Another type of non-genericity comes from renormalizability and a special R-charge assignment, which restricts the superpotential to a specific form.  The actual vacua from explicit calculation do not agree with the prediction from field counting even with generic coefficients~\cite{Sun:2019bnd}.  Such non-generic exceptions may introduce non-neglectable error to the field counting method.  Patterns of some exceptional models can be described by a sufficient condition~\cite{Amariti:2020lvx, Sun:2021svm, Li:2021ydn}, but exceptions beyond the condition are also constructed~\cite{Brister:2022rrz, Nakayama:2023eax}.

To investigate the issue of genericity, this work makes an exhaustive search for R-symmetric Wess-Zumino models with up to $5$ fields.  Models are identified by their R-charge assignments which uniquely determine the superpotentials with their coefficients taking generic values.  Redundant models related by permutation symmetries and decoupled models are properly excluded from the model data collection.  Superpotential coefficients in each model are then set to random complex values, and numerical calculation to solve the F-term equations determines whether a SUSY vacuum exists.  This procedure is repeated multiple times with new randomized coefficients to avoid non-genericity in coefficients.  The explicit results are compared with the predictions from the Nelson-Seiberg theorem.  One can estimate the accuracy of the field counting method by such a comparison.  Other information can also be extracted from the dataset of models on request.

The rest of this paper is arranged as follows.  Section \ref{sec:02} reviews the Nelson-Seiberg theorem, its extensions and exceptions.  Section \ref{sec:03} sets up the data structure to describe R-symmetric Wess-Zumino models.  Section \ref{sec:04} discusses the issue of redundancy from reducible models which should be eliminated from the dataset.  Section \ref{sec:05} presents the search algorithm and results.  Section \ref{sec:06} interpret the results and discusses other issues not covered in our current work.  Appendix \ref{app:a} gives the full list of models in our dataset.  Appendix \ref{app:b} analyzes exceptional models in the dataset.

\section{The Nelson-Seiberg theorem and exceptions}\label{sec:02}

Here we briefly summarize the Nelson-Seiberg theorem~\cite{Nelson:1993nf} and related results~\cite{Kang:2012fn, Li:2020wdk, Sun:2011fq, Brister:2022xxx, Sun:2019bnd, Amariti:2020lvx, Sun:2021svm, Li:2021ydn, Brister:2022rrz, Nakayama:2023eax}, for details we refer readers to those papers.

In the setup of a Wess-Zumino model,  the superpotential $W(z_i)$ is a holomorphic function of complex scalar fields $z_i$ or their corresponding chiral superfields.  A solution to the F-term equations
\begin{equation}\label{eq:2-01}
F_i^* = \partial_i W
      = \frac{\partial W}{\partial z_i}
      = 0.
\end{equation}
gives a SUSY vacuum, which is also a global minimum of the scalar potential
\begin{equation}
V = K^{\bar{i} j} (\partial_i W)^* \partial_j W,
\end{equation}
where $K^{\bar{i} j}$ is a the positive-definite K\"ahler metric.  A minimum of $V$ which does not solve~\eqref{eq:2-01} gives a SUSY breaking vacuum.  SUSY breaking vacua can also be constructed on runaway directions~\cite{Ferretti:2007ec, Ferretti:2007rq, Azeyanagi:2012pc, Sun:2018hnk} with corrections at large field values introduced.  By these means, the nonexistence of a solution to~\eqref{eq:2-01} can be used as a criteria for SUSY breaking.  The relation between SUSY breaking and R-symmetries under which the superpotential $W$ has R-charge $2$ is described by the Nelson-Seiberg theorem~\cite{Nelson:1993nf}:

\begin{theorem}\label{thm:01}
In a Wess-Zumino model with a generic superpotential, an R-symmetry is a necessary condition, and a spontaneously broken R-symmetry is a sufficient condition for SUSY breaking at the global minimum.
\end{theorem}

A revision of the theorem~\cite{Kang:2012fn} gives a combined necessary and sufficient condition for SUSY breaking assuming a polynomial superpotential, with a generalization to non-polynomial superpotential~\cite{Li:2020wdk}:

\begin{theorem}\label{thm:02}
In a Wess-Zumino model with a generic superpotential, the necessary and sufficient condition for SUSY breaking at the global minimum is that the superpotential has an R-symmetry, and one of the following conditions is satisfied:
\begin{itemize}
    \item The superpotential is analytic at the field space origin and $N_2 > N_0$ holds for any consistent R-charge assignment, where $N_2$ and $N_0$ are the numbers of R-charge $2$ and R-charge $0$ fields.
    \item The superpotential is non-analytic at the field space origin.
\end{itemize}
\end{theorem}

In case of polynomial superpotentials, the criteria for SUSY breaking are just counting and comparing $N_2$ and $N_0$ of each model.  The SUSY vacua from $N_2 \le N_0$ preserve the R-symmetry and give a zero expectation value to $W$~\cite{Sun:2011fq}, and this result for SUSY vacua is also true for $\mathbb{Z}_{n \ge 3}$ R-symmetries or non-Abelian discrete R-symmetries~\cite{Brister:2022xxx}.  Exceptional models with generic superpotential coefficients are also found~\cite{Sun:2019bnd} to be inconsistent with Theorem \ref{thm:01} and \ref{thm:02}.  These exceptions can be viewed as having non-generic R-charges, but are more properly described as possessing special R-charge assignments satisfying certain conditions~\cite{Amariti:2020lvx, Sun:2021svm}:

\begin{theorem}\label{thm:03}
In a Wess-Zumino model with a renormalizable superpotential and generic coefficients, a sufficient condition for the existence of a SUSY minimum is that $N_2 \le N_0 + N'_\pm$ holds for a consistent R-charge assignment, where $N_2$ and $N_0$ are the numbers of R-charge $2$ and R-charge $0$ fields, and $N'_\pm$ is the number of independent products of oppositely R-charged fields which appear only linearly in cubic terms of the superpotential.
\end{theorem}

When there is degeneracy of R-charges, we have $N'_\pm = \sum_r (N'_{+ r} + N'_{- r} - 1)$ calculated from $N'_{+ r}$ and $N'_{- r}$, the numbers of R-charge $+ r$ and $- r$ fields which appear only linearly in cubic terms of a renormalizable superpotential, and the sum is over all $r$ satisfying $r > 0$ and $r \ne 2$ with nonzero $N'_{+ r}$ and $N'_{- r}$.  This sufficient condition includes the case $N_2 \le N_0$, under which the revised Nelson-Seiberg theorem predicts the existence of SUSY vacua~\cite{Sun:2011fq}, and the case $N_0 < N_2 \le N_0 + N'_\pm$ which covers most previous exceptions~\cite{Sun:2019bnd, Amariti:2020lvx, Sun:2021svm}.  It is proved that there is no exceptional model of SUSY breaking type~\cite{Li:2021ydn}.  Thus beside generic coefficients, there is no additional genericity assumption for this sufficient condition.  This condition is not also necessary.  It is violated by recently found new exceptions~\cite{Brister:2022rrz, Nakayama:2023eax}.

The feature of generic coefficients in these exceptions implies that they may introduce non-negligible error to the field counting method based on Theorem \ref{thm:02} or \ref{thm:03}.  A straightforward way to check the magnitude of error is to collect a dataset of R-symmetric Wess-Zumino models, compare their vacuum solutions with the predictions from the Nelson-Seiberg theorem, and give the proportion of inconsistency in all the models.  In the following sections we present a preliminary construction of such a dataset.

\section{The data structure of models}\label{sec:03}

Our goal is to exhaustively search R-symmetric Wess-Zumino models in a reasonable scope, e.g.\ with $N$ fields
\begin{equation}
\hat z = \{ z_i \}
       = \{ z_1, \dots , z_N \}.
\end{equation}
We consider models with their R-charge assignments fixed, or equivalently, models without extra $U(1)$ symmetries which are not R-symmetries~\cite{Komargodski:2009jf}.  Each model is uniquely labeled by its R-charge assignment, recorded as an array of R-charges:
\begin{equation}
\hat r = \{ r_i \}
       = \{ r_1, \dots , r_N \}
       = \{ R(z_1), \dots , R(z_N) \}.
\end{equation}
A renormalizable superpotential is a cubic polynomial function of $\hat z$:
\begin{equation}
W = c + c_i z_i + c_{i j} z_i z_j + c_{i j k} z_i z_j z_k.
\end{equation}
In a generic R-symmetric model, $c$, $c_i$, $c_{i j}$ and $c_{i j k}$ are generic coefficients for R-charge 2 terms, and zero for terms with other R-charges.  It is obvious that the R-symmetry sets $c = 0$.  We rewrite $W$ in a compact notation:
\begin{equation}\label{eq:3-01}
W = \hat c \cdot \hat z^{\hat a}
  = c_s \hat z^{\hat a^s}
  = c_s z_1^{a^s_1} \dots z_N^{a^s_N}
  = c_1 z_1^{a^1_1} \dots z_N^{a^1_N} + \dots + c_m z_1^{a^m_1} \dots z_N^{a^m_N},
\end{equation}
where the sum over $s$ only includes terms respecting the R-symmetry and with non-zero coefficients
\begin{equation}
\hat c = \{ c_s \}
       = \{ c_1, \dots , c_m \}.
\end{equation}
Because of genericity in coefficients, we identify $W$ of the same polynomial form with different coefficient values as the same model.  Thus each model is also uniquely labeled by the two-dimensional array of exponents of all fields in all monomial terms of $W$:
\begin{equation}
\hat a = \{ \hat a^s \}
       = \{ a^s_i \}
        = \{ \hat a^1, \dots , \hat a^m \}
        = \{ \{ a^1_1, \dots , a^1_N \},
             \dots ,
             \{ a^m_1, \dots , a^m_N \} \},
\end{equation}
subject to the renormalization constraint
\begin{equation}\label{eq:3-02}
a^s_1 + \dots + a^s_N \in \{ 0, 1, 2, 3 \}, \quad
a^s_i \in \{ 0, 1, 2, 3 \}.
\end{equation}
and the R-symmetry constraint
\begin{equation}\label{eq:3-03}
\hat a^s \cdot \hat r = a^s_i r_i
                      = a^s_1 r_1 + \dots + a^s_N r_N
                      = 2,
\end{equation}
Which uniquely fixes the array of R-charges $\hat r$.  On the other hand, given an array of R-charges $\hat r$, the array of exponents $\hat a$ can be uniquely determined by including all $\hat a^s$ satisfying the constraints \eqref{eq:3-02} and \eqref{eq:3-03}.

To construct the polynomial form of $W$ or its corresponding array of exponents $\hat a$, we first collect all possible monomials of $\hat z$ up to cubic, or the two-dimensional array of exponents of all fields in all possible monomials:
\begin{equation}
\hat A = \{ \hat A^I \}
       = \{ A^I_i \}
       = \{ \hat A^1, \dots , \hat A^M \}
       = \{ \{ A^1_1, \dots , A^1_N \},
            \dots ,
            \{ A^M_1, \dots , A^M_N \} \},
\end{equation}
subject to the renormalization constraint
\begin{equation}\label{eq:3-04}
A^I_1 + \dots + A^I_N \in \{ 0, 1, 2, 3 \}, \quad
A^I_i \in \{ 0, 1, 2, 3 \}.
\end{equation}
Each R-symmetric Wess-Zumino model corresponds to an array of indices
\begin{equation}
\hat I = \{ I_1, \dots , I_m \}.
\end{equation}
It selects $m$ subarray of exponents from $\hat A$:
\begin{equation}\label{eq:3-05}
\hat a = \{ \hat a^s \}
       = \{ \hat A^{I_s} \}
\end{equation}
safisfying the R-symmetry constraint
\begin{equation}\label{eq:3-06}
\hat A^{I_s} \cdot \hat r = A^{I_s}_i r_i
                          = A^{I_s}_1 r_1 + \dots + A^{I_s}_N r_N
                          = 2.
\end{equation}
Therefore models can be equivalently labeled with $\hat I$ if $\hat A$ has been recorded at first, and the polynomial form of $W$ can be constructed through \eqref{eq:3-01} and \eqref{eq:3-05}.  Note that $m$, the number of non-zero terms in $W$, can vary in different models, while $M$, the number of all possible monomials up to cubic, is solely determined by $N$.

In the dataset, we record each model as its corresponding array of R-charges $\hat r \in \{ \hat r^{(p)} \}$.  R-charges $r_i$ in $\hat r$ are sorted so that R-charges $2$, $0$ and $- 2$ come first, followed by other R-charges in ascending order of their absolute values, with the positive R-charges preceding the negative R-charges if their absolute values are the same, i.e.:
\begin{equation}\label{eq:3-07}
\begin{split}
i < j \ \Leftrightarrow \ r_i \preceq r_j
      \ &\Leftrightarrow \ (r_i = r_j)
                           \lor (r_i = 2)
                           \lor (r_i = 0 \land r_j \ne 2)\\
        &\qquad
                           \lor (r_i = - 2 \land r_j \ne 2 \land r_j \ne 0)\\
        &\qquad
                           \lor (\lvert r_i \rvert < \lvert r_j \rvert \ne 2)
                           \lor (r_i = - r_j > 0).
\end{split}
\end{equation}
All possible monomials of $\hat z$ up to cubic are recorded as the array of exponents $\hat A$.  Monomials or subarrays of exponents $\hat A^I$ in $\hat A$ are sorted in ascending order of their values represented in little-endian or the least significant digit first (LSDF) convention, i.e.:
\begin{equation}\label{eq:3-08}
I < J \ \Leftrightarrow \ \hat A^I \prec \hat A^J
      \ \Leftrightarrow \ (A^I_i = A^J_i \ \text{for} \ i = k + 1, \dots , N)
                          \land (A^I_k < A^J_k).
\end{equation}
The array of indices $\hat I$ can be obtained from $\hat r$ by including all $I_s$ whose corresponding $\hat{A}^{I_s}$ satisfy \eqref{eq:3-06}.  Indices $I_s$ in $\hat I$ are sorted in ascending order:
\begin{equation}\label{eq:3-09}
s < t \ \Leftrightarrow \ I_s < I_t.
\end{equation}
Finally, model records $\hat r^{(p)}$ in the dataset $\{ \hat r^{(p)} \}$ are sorted in ascending order of their values represented in big-endian or the most significant digit first (MSDF) convention with respect to the individual R-charge ordering \eqref{eq:3-07}, i.e.:
\begin{equation}\label{eq:3-10}
p < q \ \Leftrightarrow \ \hat r^{(p)} \prec \hat r^{(q)}
      \ \Leftrightarrow \ (r^{(p)}_i = r^{(q)}_i \ \text{for} \ i = 1, \dots , k - 1)
                          \land (r^{(p)}_k \prec r^{(q)}_k).
\end{equation}

As an example, there are in total $M = 10$ possible monomials up to cubic for $N = 2$:
\begin{equation}
\{ 1 \text{(a constant term)}, z_1, z_1^2, z_1^3,
   z_2, z_1 z_2, z_1^2 z_2,
   z_2^2, z_1 z_2^2, z_2^3 \}.
\end{equation}
They are recorded as
\begin{equation}
\begin{split}
\hat A &= \{ \{ A^I_1, A^I_2 \} \mid I = 1, \dots , 10 \}\\
       &= \{ \{ 0, 0 \}, \{ 1, 0 \}, \{ 2, 0 \}, \{ 3, 0 \},
             \{ 0, 1 \}, \{ 1, 1 \}, \{ 2, 1 \},
             \{ 0, 2 \}, \{ 1, 2 \}, \{ 0, 3 \} \},
\end{split}
\end{equation}
The deformed Polonyi model~\cite{Intriligator:2007cp} is recorded as
\begin{equation}
\hat r = \{ r_1, r_2 \}
       = \{ 2, 0 \}.
\end{equation}
The R-symmetry constraint \eqref{eq:3-06} gives
\begin{align}
\hat I &= \{ I_1, I_2, I_3 \}
        = \{ 2, 6, 9 \},\\
\hat a &= \{ \hat a^1, \hat a^2, \hat a^3 \}
        = \{ \hat A^{I_1}, \hat A^{I_2}, \hat A^{I_3} \}
        = \{ \hat A^2, \hat A^6, \hat A^9 \}
        = \{ \{ 1, 0 \}, \{ 1, 1 \}, \{ 1, 2 \} \}.
\end{align}
Thus the superpotential is constructed with generic coefficients:
\begin{equation}
W = c_s \hat z^{\hat a^s}
  = c_1 z_1 + c_2 z_1 z_2 + c_3 z_1 z_2^2.
\end{equation}
Note that the redundancy from permutation of superpotential terms such as
\begin{equation}
W = c_2 z_1 z_2 + c_3 z_1 z_2^2 + c_1 z_1.
\ \Leftrightarrow \ \hat I = \{ 6, 9, 2 \}
\ \Leftrightarrow \ \hat r = \{ 2, 0 \}
\end{equation}
and from permutation of fields such as
\begin{equation}
W = c_1 z_2 + c_2 z_1 z_2 + c_3 z_1^2 z_2
\ \Leftrightarrow \ \hat I = \{ 5, 6, 7 \}
\ \Leftrightarrow \ \hat r = \{ 0, 2 \}
\end{equation}
has been eliminated by sorting the data in the order of \eqref{eq:3-07}, \eqref{eq:3-08} and \eqref{eq:3-09}.

\section{Redundancy from reducible models}\label{sec:04}

Another type of redundancy comes from models which can be reduced to two or more decoupled subsystems recorded in the dataset as models with fewer fields.  The array of exponents of a reducible model has the reduction:
\begin{equation}\label{eq:4-01}
\begin{gathered}
\hat a = \hat a' \oplus \hat a''
       = \{ \hat a^{s_1}, \dots , \hat a^{s_l} \}
             \oplus \{ \hat a^{s_{l + 1}}, \dots , \hat a^{s_m} \},\\
\hat a^{s_u} \cdot \hat a^{s_{v}} = \hat a^{s_u}_i \hat a^{s_{v}}_i
                                  = 0 \quad
\text{for} \quad
u \in \{ 1, \dots , l \}, \
v \in \{ l + 1, \dots , m \}, \\
\text{for a permutation} \quad
p^{(s)} = \begin{pmatrix}
          1 & \dots & m\\
          s_1 & \dots & s_m
          \end{pmatrix}
        \in S_m \quad
\text{and} \quad
l \in \{ 1, \dots , m - 1 \}.
\end{gathered}
\end{equation}
This condition can be interpreted as reducing $\hat a$ into two parts orthogonal to each other.  The array of fields and their R-charges also have the reduction
\begin{equation}
\begin{gathered}
\hat z = \hat z' \oplus \hat z''
        = \{ z_{i_1}, \dots , z_{i_L}\} \oplus \{ z_{i_{L + 1}}, \dots , z_{i_N} \},\\
\hat r = \hat r' \oplus \hat r''
        = \{ r_{i_1}, \dots , r_{i_L}\} \oplus \{ r_{i_{L + 1}}, \dots , r_{i_N} \},\\
\text{for a permutation} \quad
p^{(i)} = \begin{pmatrix}
          1 & \dots & N\\
          i_1 & \dots & i_N
          \end{pmatrix}
        \in S_N \quad
\text{and} \quad
L \in \{ 1, \dots , N - 1 \}.     
\end{gathered}
\end{equation}
With such permutation of superpotential terms $p^{(s)}$ and permutation of fields $p^{(i)}$, the array of exponents $\hat a$ is block diagonalized:
\begin{equation}
\begin{gathered}
a^{s_u}_{i_k} = 0,\\
\text{for} \quad
(u \in \{ 1, \dots , l \}, k \in \{ L+1, \dots , N \})
\lor (u \in \{ l+1, \dots , m \}, k \in \{ 1, \dots , L \}).
\end{gathered}
\end{equation}

As an example, the model with four fields
\begin{equation}
\begin{gathered}
\hat r = \{ 1 / 2, 2 / 3, 1, 4 / 3 \}, \quad
\hat a = \{ \{ 0, 3, 0, 0 \}, \{ 2, 0, 1, 0 \}, \{ 0, 0, 2, 0 \}, \{ 0, 1, 0, 1 \} \},\\
W = c_1 z_2^3 + c_2 z_1^2 z_3 + c_3 z_3^2 + c_4 z_2 z_4
\end{gathered}
\end{equation}
has the reduction into two models:
\begin{equation}
\begin{gathered}
\hat r = \hat r' \oplus \hat r''
        = \{ 1 / 2 , 1 \} \oplus \{ 2 / 3, 4 / 3\},\\
\hat a = \hat a' \oplus \hat a''
        = \{ \{ 2, 1 \}, \{ 0, 2 \} \} \oplus \{ \{ 3, 0 \}, \{ 1, 1 \} \},\\
W = W' + W''
  = (c_2 z_1^2 z_3 + c_3 z_3^2) + (c_1 z_2^3 + c_4 z_2 z_4).
\end{gathered}
\end{equation}
If $\hat r'$ and $\hat r''$ are already in the dataset of models with two fields, recording $\hat r$ as a model with four fields becomes redundant, because it is equivalent to two decoupled sectors with $\hat r'$ and $\hat r''$.  To eliminate such redundancy, it is necessary and sufficient to exclude models satisfying the condition \eqref{eq:4-01}.  Therefore our dataset only contains records of irreducible models.

\section{The search algorithm and results}\label{sec:05}

Equipped with the data structure introduced in previous sections, we are ready to construct the dataset of R-symmetric Wess-Zumino models.  Here we give a brief description of the brute-force search algorithm:

\begin{enumerate}
\item Generate $\hat A$, the array of exponents of all fields in all possible monomials.\label{st:1}
      \begin{enumerate}
      \item Fix $N$ and initialize $\hat A$ to an empty array.
      \item For each ordered $N$-tuple of $\{ 0, 1, 2, 3 \}$, append the tuple into $\hat A = \{ \hat A^I \}$ if the tuple satisfies the renormalization constraint \eqref{eq:3-04}.
      \item Let $M$ be the length of $\hat A$ after the previous step, i.e., $I \in \{ 1, \dots , M \}$.
      \item Sort the elements $\hat A^I$ of $\hat A$ in the order of \eqref{eq:3-08}.
      \end{enumerate}
\item Generate $\{ \hat r^{(p)} \}$, the dataset of models recorded  as their R-charge assignments.
      \begin{enumerate}
      \item Initialize $\{ \hat r^{(p)} \}$ to an empty array.
      \item For each unordered $N$-tuple of $\{ 1, \dots, M \}$, do the following steps for the tuple or array $\hat I = \{ I_s \}$:\label{st:2b}
            \begin{enumerate}
            \item Try to solve the R-symmetry constraint \eqref{eq:3-03} where $\hat a = \{ \hat a^s \} =\{ \hat A^{I_s} \}$ is determined by $\hat I$, and continue the following steps if a unique solution $\hat r$ exists.\label{st:2bi}
            \item Sort the elements $r_i$ of $\hat r$ in the order of \eqref{eq:3-07};
            \item For each previously recorded array of R-charges $\hat r^{(p)}$, compare $\hat r$ with $\hat r^{(p)}$, and continue the following steps if $\hat r$ is not previously recorded in $\{ \hat r^{(p)} \}$.
            \item Reset $\hat a$ to empty arrays.\label{st:2biv}
            \item For each $J \in \{ 1, \dots, M \}$, append $\hat A^J$ to $\hat a$ if $\hat A^J$ satisfies the R-symmetry constraints \eqref{eq:3-06}.
            \item Let $m$ be the length of $\hat a = \{ \hat a^s \}$ after the previous step, i.e., $s \in \{ 1, \dots , m \}$.
            \item For each permutation $p^{(s)} \in S_m$ and $l \in \{ 1, \dots , m \}$, check the reduction condition \eqref{eq:4-01}, and append $\hat r$ into $\{ \hat r^{(p)} \}$ if the model is verified to be irreducible.\label{st:2bvii}
            \end{enumerate}
      \item Let $L$ be the length of $\{ \hat r^{(p)} \}$ after the previous step, i.e., $p \in \{ 1, \dots , L \}$
      \item Sort the elements $\hat r^{(p)}$ of $\{ \hat r^{(p)} \}$ in the order of \eqref{eq:3-10}.
      \end{enumerate}
\item Classify models according to their R-charge assignments and whether they have SUSY solutions or not.  For each $p \in \{ 1, \dots , L \}$, do the following steps for the record $\hat r = \hat r^{(p)}$:\label{st:3}
      \begin{enumerate}
      \item Count $2$'s and $0$'s in $\hat r$, and record whether the model belongs to the $N_2 \le N_0$ type or the $N_2 > N_0$ type.
      \item Initialize $\hat a$ to an empty array.
      \item For each $J \in \{ 1, \dots, M \}$, append $\hat A^J$ to $\hat a$ if $\hat A^J$ satisfies the R-symmetry constraints \eqref{eq:3-06} with the R-charge assignment $\hat r$.
      \item Construct the polynomial form of the superpotential $W$ from $\hat a$ through \eqref{eq:3-01}.
      \item Repeat the following procedure a number of times: generate random coefficients $\hat c$ for $W$ from the previous step, try to numerically solve the F-term equations \eqref{eq:2-01}, and record whether a SUSY solution exists or not.\label{st:3e}
      \end{enumerate}
\end{enumerate}

In step \ref{st:2bi}, we have used the fact that $N$ linear equations in $N$ variables with a full-rank coefficient matrix form a balanced system which has a unique solution.  On the other hand, if a model with $N$ fields uniquely fixes $\hat r$, it must be possible to select $N$ monomial terms from $W$ with their array of exponents $\hat a$ being full rank, and the R-symmetry constraint \eqref{eq:3-03} with $\hat a$ has $\hat r$ as the unique solution.  Therefore the exhaustion of all unordered $N$-tulpes of $\{ 1, \dots, M \}$, or all combinations of $N$ monomial up to cubic, gives all possible values of $\hat r$ for R-symmetric Wess-Zumino models with $N$ fields, and the dataset constructed from this brute-force search algorithm is complete.

In step \ref{st:3}, for each record $\hat r^{(p)}$ in the dataset constructed in step \ref{st:2b}, we classify the model by field counting through the revised Nelson-Seiberg theorem, and make a numerical calculation to explicitly check the existence of a SUSY solution.  To implement the genericity assumption, the numerical calculation is repeated multiple times with different random values for superpotential coefficients.  Comparing the explicit result with the prediction from the field counting method, one can identify whether the model agrees or violates the Nelson-Seiberg theorem.

Following the algorithm described here, our current code has constructed the dataset for $N \le 5$.  The search result is summarized in Table \ref{tb:01}.  The full list of models is presented in Appendix \ref{app:a}.  In total there are $859$ R-symmetric Wess-Zumino models with $N \le 5$ fields.  $556$ models have $N_2 \le N_0$ and SUSY solutions, and $284$ models have $N_2 > N_0$ and no SUSY solution.  These two types of models are in agreement with the Nelson-Seiberg theorem.  $19$ models in the dataset have $N_2 > N_0$ and SUSY solutions.  Explicit calculation also shows that R-symmetries are spontaneously broken at SUSY vacua in these models.  Thus the last type of models are non-generic exceptions to both Theorem \ref{thm:01} and \ref{thm:02}.  These exceptional models are analyzed in Appendix \ref{app:b}.  Most of them except two have their R-charge assignments satisfying the condition in Theorem \ref{thm:03}.  The two exceptions, with R-charge assignments $\hat r = \{ 2, - 2, 4, - 4, 8 \}$ and $\hat r = \{2, \frac{2}{3}, \frac{4}{3}, -\frac{4}{3}, \frac{8}{3} \}$, fall within the scope of~\cite{Nakayama:2023eax}.
 
\begin{table}
\centering
\renewcommand\arraystretch{1.1}
\begin{tabular}{*{7}{c}}
\hline
\multirow{2}*{Model Type} & \multicolumn{6}{c}{Number of Models}\\
\cline{2-7}
                          & $N = 1$ & $N = 2$ & $N = 3$ & $N = 4$ & $N = 5$ & Total\\
\hline
$N_2 \le N_0$, SUSY       & $2$     & $6$     & $19$    & $81$    & $448$   & $556$\\
$N_2 > N_0$, no SUSY      & $1$     & $1$     & $7$     & $38$    & $237$   & $284$\\
$N_2 > N_0$, SUSY         & $0$     & $0$     & $0$     & $1$     & $18$    & $19$\\
Total                     & $3$     & $7$     & $26$    & $120$   & $703$   & $859$\\
\hline
\end{tabular}
\caption{Summary of R-symmetric Wess-Zumino models with $N \le 5$ fields.}\label{tb:01}
\end{table}

\section{Interpretation and Discussion}\label{sec:06}

The identification of ``SUSY'' or ``no SUSY'' type in Table \ref{tb:01} is according to numerical search for a solution to the F-term equations \eqref{eq:2-01} with random superpotential coefficients.  All models including the $19$ exceptions have given consistent results with different coefficient values in each loop of step \ref{st:3e}.  Therefore we are very sure that the results in Table \ref{tb:01} are based on the assumption of generic coefficients.  The occurrence of exceptions in our dataset is $19$ out of $859$ in total.  The number $19 / 859 \approx 2.2 \%$, which is small but not negligible, can be used as an estimation of the failure rate of the field counting method based on the revised Nelson-Seiberg theorem.

The SUSY vacua in these $19$ exceptions give $W = 0$ like any R-symmetric Wess-Zumino model possessing a SUSY vacuum~\cite{Kappl:2008ie, Dine:2009sw, Brister:2021xca}.  SUGRA extensions of these models also give SUSY vacua with $V = 0$, and contribute to the low-scale SUSY branch of the landscape.  As stated in~\cite{Li:2021ydn}, there is no generic model with $N_2 \le N_0$ and SUSY breaking.  Therefore if we use the field-counting method to classify out models with SUSY vacua, all exceptions are false negatives, or type II errors in binary classification, which affect the recall but retain the precision of the classifier.

Furthermore, when SUGRA extensions of Wess-Zumino model are used as the low energy description for flux compactification of type IIB superstring theory, the R-symmetry breaking property of SUSY vacua in exceptional models means that the moduli values are away from the R-symmetric point of the moduli space.  Then it is unnatural to turn on only R-symmetric fluxes and obtain an R-symmetric effective superpotential if the configuration of the Calabi-Yau manifold already breaks the R-symmetry.  Therefore these exceptions introduce no error to the field counting method if we only consider R-symmetric SUSY vacua in the low-energy SUSY branch~\cite{Dine:2005gz, Dine:2004is, Dine:2005yq}, or string vacua with enhanced symmetries~\cite{DeWolfe:2004ns, DeWolfe:2005gy, Palti:2007pm, Kanno:2017nub, Palti:2020qlc, Kanno:2020kxr}.  Whether the R-symmetry breaking SUSY vacua have other natural realizations in string models remains an open question.

The dataset may include more R-symmetric Wess-Zumino models if we relax the renormalization requirement, and more exceptions may appear.  For example, if we allow $W$ to be up to quartic, the model with
\begin{equation}
\hat r = \{ r_1, r_2, r_3 \}
       = \{ 2, 4, -4 \}
\end{equation}
gives the generic superpotential
\begin{equation}
W = c_1 z_1 + c_2 z_1^3 z_3 + c_3 z_1 z_2 z_3.
\end{equation}
The model has $N_2 > N_0$ but gives a class of degenerate SUSY vacua at
\begin{equation}
z_1 = 0, \quad
z_2 z_3 = - c_1 / c_3.
\end{equation}
It is an exception to both Theorem \ref{thm:01} and \ref{thm:02} with only $N = 3$ fields, which is not in our dataset.  If the superpotential is required to be renormalizable, the search result in Table \ref{tb:01} shows that exceptions only appear from $N = 4$ on, and the simplest exception with $N = 4$ fields has been presented before in literature~\cite{Sun:2019bnd}.

Even within the scope of renormalizability, there are still unexplored problems in our dataset.  We have searched models with all R-charges fixed, thus models with ambiguity in the R-charge assignment, or equivalently with an extra $U(1)$ symmetry~\cite{Komargodski:2009jf}, are excluded.  The restriction from extra symmetries may introduce models with new features.  Another issue is that we use the F-term equations $\partial_i W = 0$ to identify the model as of ``SUSY'' or ``no SUSY'' type.  If our goal is just to classify out models with SUSY vacua, both metastable vacua and runaway directions affect neither the recall nor the precision of the classifier.  But these features in vacuum structure need to be studied if the dataset is used for other purposes.

The time complexity of our algorithm can be estimated as follows.  For a certain number of fields $N$, the length of $\hat A$ or the number of monomials up to cubic is
\begin{equation}
M = \frac{1}{6} (N+1) (N+2) (N+3)
  = O(N^3),
\end{equation}
which is also the minimal time complexity of step \ref{st:1} of the search algorithm.  The outermost loop of step \ref{st:2b} goes through all combinations of $N$ monomials from all the $M$ monomials obtained in step \ref{st:1}.  The number of combinations is equal to the binomial coefficient
\begin{equation}\label{eq:6-01}
C = \binom{M}{N}
  = \frac{M!}{N! (M - N)!}
  = O(N^{2 N}),
\end{equation}
where Stirling's approximation is used to derive the last equality.  All the inner loops of step \ref{st:2b} have polynomial time complexity with properly optimized algorithms.  So the total time complexity of the dataset construction is $O(N^{2 N})$, and the algorithm belongs to the complexity class with exponential run time (EXP).  On the other hand, whether a given $\hat r$ is in the dataset can be verified in polynomial run time through the procedure from step \ref{st:2biv} to step \ref{st:2bvii} to exclude reducible models, as well as step \ref{st:2bi} to ensure the uniqueness of R-charges.  Thus finding one solution $\hat{r}$ belongs to the complexity class with non-deterministic polynomial time (NP).  But constructing the whole model dataset is more complicated and can not be classified as NP.

The brute-force search of the dataset with a larger $N$ requires more run time which grows exponentially.  Using the estimation~\eqref{eq:6-01}, the time complexities for $N = 6$ and $N = 7$ are respectively about $200$ and $70000$ times of the time complexity for $N = 5$.  A better search algorithm and optimized code may reduce the run time, but unless there is an algorithm of polynomial time complexity, the exhaustive search quickly becomes impracticable as $N$ grows, and one must turn to a non-exhaustive sampling.  Eventually, if SUGRA extensions of Wess-Zumino models are used as the low energy description for flux compactification of type IIB superstring theory, the number of fields $N$ corresponds to the number of the complex structure moduli, or the Hodge number $h^{2, 1}$ of the Calabi-Yau manifold, which can be up to several hundred for some known Calabi-Yau manifolds~\cite{Candelas:1987kf, Kreuzer:1992bi, Kreuzer:1992np, Avram:1996pj, Kreuzer:2000qv, Kreuzer:2000xy, Johnson:2014xpa, Anderson:2015iia, Anderson:2017aux, CYdata:websites}.  We expect that new technologies such as machine learning~\cite{Ruehle:2020jrk} may help us to construct the dataset for such a huge $N$.

\section*{Acknowledgement}

The authors thank Xin Gao, Jinmian Li, Tianjun Li, Yu Nakayama and Greg Yang for helpful discussions.  This work is supported by the National Natural Science Foundation of China under the grant number 12205208 and 11305110.

\appendix

\section{List of models}\label{app:a}

The search result includes generic R-symmetric Wess-Zumino models with $N \le 5$ fields, recorded as their R-charge assignments $\hat r$ in the list.  Models are classified according to whether they belong to the $N_2 \le N_0$ type or the $N_2 > N_0$ type and whether they have SUSY solutions or not.

\subsection{Models with $N = 1$}

There are $3$ models in total, $2$ of them have $N_2 \le N_0$ and SUSY solutions:
\begin{equation}
\hat r = \{ \tfrac{2}{3} \}, \
         \{ 1 \};
\end{equation}
and $1$ of them has $N_2 > N_0$ and no SUSY solution:
\begin{equation}
\hat r = \{ 2 \}.
\end{equation}

\subsection{Models with $N = 2$}

There are $7$ models in total, $6$ of them have $N_2 \le N_0$ and SUSY solutions:
\begin{equation}
\hat r = \{ 2, 0 \}, \
         \{ 0, 1 \}, \
         \{ \tfrac{1}{2}, 1 \}, \
         \{ \tfrac{2}{3}, \tfrac{2}{3} \}, \
         \{ \tfrac{2}{3}, \tfrac{4}{3} \}, \
         \{ 1, 1 \};
\end{equation}
and $1$ of them has $N_2 > N_0$ and no SUSY solution:
\begin{equation}
\hat r = \{ 2, -2 \}.
\end{equation}

\subsection{Models with $N = 3$}

There are $26$ models in total, $19$ of them have $N_2 \le N_0$ and SUSY solutions:
\begin{align*}
\hat r = \mbox{}
         &\{ 2, 0, 0 \},&
         &\{ 2, 0, -2 \},&
         &\{ 2, 0, 1 \},&
         &\{ 0, 0, 1 \},&
         &\{ 0, \tfrac{1}{2}, 1 \},&
         &\{ 0, \tfrac{2}{3}, \tfrac{4}{3} \},&
         &\{ 0, 1, 1 \},\\
         &\{ \tfrac{1}{3}, \tfrac{2}{3}, 1 \},&
         &\{ \tfrac{1}{3}, \tfrac{2}{3}, \tfrac{4}{3} \},&
         &\{ \tfrac{2}{5}, \tfrac{4}{5}, \tfrac{6}{5} \},&
         &\{ \tfrac{1}{2}, \tfrac{1}{2}, 1 \},&
         &\{ \tfrac{1}{2}, \tfrac{3}{4}, 1 \},&
         &\{ \tfrac{1}{2}, 1, 1 \},&
         &\{ \tfrac{1}{2}, 1, \tfrac{3}{2} \},\\
         &\{ \tfrac{2}{3}, \tfrac{2}{3}, \tfrac{2}{3} \},&
         &\{ \tfrac{2}{3}, \tfrac{2}{3}, \tfrac{4}{3} \},&
         &\{ \tfrac{2}{3}, -\tfrac{2}{3}, \tfrac{4}{3} \},&
         &\{ \tfrac{2}{3}, \tfrac{4}{3}, \tfrac{4}{3} \},&
         &\{ 1, 1, 1 \}; \refstepcounter{equation} \tag{\theequation}
\end{align*}
and $7$ of them have $N_2 > N_0$ and no SUSY solution:
\begin{equation}
\hat r = \{ 2, 2, 0 \}, \
         \{ 2, 2, -2 \}, \
         \{ 2, -2, -2 \}, \
         \{ 2, -2, 4 \}, \
         \{ 2, -2, 6 \}, \
         \{ 2, \tfrac{2}{3}, -\tfrac{2}{3} \}, \
         \{ 2, 1, -1 \}.
\end{equation}

\subsection{Models with $N = 4$}

There are $120$ models in total, $81$ of them have $N_2 \le N_0$ and SUSY solutions:
\begin{align*}
\hat r = \mbox{}
         &\{ 2, 2, 0, 0 \},&
         &\{ 2, 0, 0, 0 \},&
         &\{ 2, 0, 0, -2 \},&
         &\{ 2, 0, 0, 1 \},&
         &\{ 2, 0, -2, -2 \},\\
         &\{ 2, 0, -2, 1 \},&
         &\{ 2, 0, -2, 4 \},&
         &\{ 2, 0, -2, 6 \},&
         &\{ 2, 0, \tfrac{1}{2}, 1 \},&
         &\{ 2, 0, \tfrac{2}{3}, -\tfrac{2}{3} \},\\
         &\{ 2, 0, \tfrac{2}{3}, \tfrac{4}{3} \},&
         &\{ 2, 0, 1, 1 \},&
         &\{ 2, 0, 1, -1 \},&
         &\{ 0, 0, 0, 1 \},&
         &\{ 0, 0, \tfrac{1}{2}, 1 \},\\
         &\{ 0, 0, \tfrac{2}{3}, \tfrac{4}{3} \},&
         &\{ 0, 0, 1, 1 \},&
         &\{ 0, \tfrac{1}{3}, \tfrac{2}{3}, 1 \},&
         &\{ 0, \tfrac{1}{3}, \tfrac{2}{3}, \tfrac{4}{3} \},&
         &\{ 0, \tfrac{2}{5}, \tfrac{4}{5}, \tfrac{6}{5} \},\\
         &\{ 0, \tfrac{1}{2}, \tfrac{1}{2}, 1 \},&
         &\{ 0, \tfrac{1}{2}, \tfrac{3}{4}, 1 \},&
         &\{ 0, \tfrac{1}{2}, 1, 1 \},&
         &\{ 0, \tfrac{1}{2}, 1, \tfrac{3}{2} \},&
         &\{ 0, \tfrac{2}{3}, \tfrac{2}{3}, \tfrac{4}{3} \},\\
         &\{ 0, \tfrac{2}{3}, -\tfrac{2}{3}, \tfrac{4}{3} \},&
         &\{ 0, \tfrac{2}{3}, 1, \tfrac{4}{3} \},&
         &\{ 0, \tfrac{2}{3}, \tfrac{4}{3}, \tfrac{4}{3} \},&
         &\{ 0, 1, 1, 1 \},&
         &\{ \tfrac{2}{9}, \tfrac{2}{3}, \tfrac{8}{9}, \tfrac{10}{9} \},\\
         &\{ \tfrac{1}{4}, \tfrac{1}{2}, \tfrac{3}{4}, 1 \},&
         &\{ \tfrac{1}{4}, \tfrac{1}{2}, 1, \tfrac{3}{2} \},&
         &\{ \tfrac{2}{7}, \tfrac{4}{7}, \tfrac{6}{7}, \tfrac{8}{7} \},&
         &\{ \tfrac{2}{7}, \tfrac{4}{7}, \tfrac{6}{7}, \tfrac{10}{7} \},&
         &\{ \tfrac{1}{3}, \tfrac{1}{3}, \tfrac{2}{3}, 1 \},\\
         &\{ \tfrac{1}{3}, \tfrac{1}{3}, \tfrac{2}{3}, \tfrac{4}{3} \},&
         &\{ \tfrac{1}{3}, \tfrac{1}{2}, \tfrac{2}{3}, 1 \},&
         &\{ \tfrac{1}{3}, \tfrac{2}{3}, \tfrac{2}{3}, 1 \},&
         &\{ \tfrac{1}{3}, \tfrac{2}{3}, \tfrac{2}{3}, \tfrac{4}{3} \},&
         &\{ \tfrac{1}{3}, \tfrac{2}{3}, -\tfrac{2}{3}, \tfrac{4}{3} \},\\
         &\{ \tfrac{1}{3}, \tfrac{2}{3}, \tfrac{5}{6}, 1 \},&
         &\{ \tfrac{1}{3}, \tfrac{2}{3}, \tfrac{5}{6}, \tfrac{4}{3} \},&
         &\{ \tfrac{1}{3}, \tfrac{2}{3}, 1, 1 \},&
         &\{ \tfrac{1}{3}, \tfrac{2}{3}, 1, \tfrac{4}{3} \},&
         &\{ \tfrac{1}{3}, \tfrac{2}{3}, 1, \tfrac{5}{3} \},\\
         &\{ \tfrac{1}{3}, \tfrac{2}{3}, \tfrac{4}{3}, \tfrac{4}{3} \},&
         &\{ \tfrac{1}{3}, \tfrac{2}{3}, \tfrac{4}{3}, \tfrac{5}{3} \},&
         &\{ -\tfrac{1}{3}, \tfrac{2}{3}, 1, \tfrac{4}{3} \},&
         &\{ \tfrac{2}{5}, \tfrac{2}{5}, \tfrac{4}{5}, \tfrac{6}{5} \},&
         &\{ \tfrac{2}{5}, -\tfrac{2}{5}, \tfrac{4}{5}, \tfrac{6}{5} \},\\
         &\{ \tfrac{2}{5}, \tfrac{3}{5}, \tfrac{4}{5}, 1 \},&
         &\{ \tfrac{2}{5}, \tfrac{3}{5}, \tfrac{4}{5}, \tfrac{6}{5} \},&
         &\{ \tfrac{2}{5}, \tfrac{4}{5}, \tfrac{4}{5}, \tfrac{6}{5} \},&
         &\{ \tfrac{2}{5}, \tfrac{4}{5}, \tfrac{6}{5}, \tfrac{6}{5} \},&
         &\{ \tfrac{2}{5}, \tfrac{4}{5}, \tfrac{6}{5}, \tfrac{8}{5} \},\\
         &\{ \tfrac{4}{9}, \tfrac{2}{3}, \tfrac{8}{9}, \tfrac{10}{9} \},&
         &\{ \tfrac{1}{2}, \tfrac{1}{2}, \tfrac{1}{2}, 1 \},&
         &\{ \tfrac{1}{2}, \tfrac{1}{2}, \tfrac{3}{4}, 1 \},&
         &\{ \tfrac{1}{2}, \tfrac{1}{2}, 1, 1 \},&
         &\{ \tfrac{1}{2}, \tfrac{1}{2}, 1, \tfrac{3}{2} \},\\
         &\{ \tfrac{1}{2}, -\tfrac{1}{2}, 1, \tfrac{3}{2} \},&
         &\{ \tfrac{1}{2}, \tfrac{5}{8}, \tfrac{3}{4}, 1 \},&
         &\{ \tfrac{1}{2}, \tfrac{2}{3}, \tfrac{5}{6}, 1 \},&
         &\{ \tfrac{1}{2}, \tfrac{3}{4}, \tfrac{3}{4}, 1 \},&
         &\{ \tfrac{1}{2}, \tfrac{3}{4}, 1, 1 \},\\
         &\{ \tfrac{1}{2}, \tfrac{3}{4}, 1, \tfrac{5}{4} \},&
         &\{ \tfrac{1}{2}, \tfrac{3}{4}, 1, \tfrac{3}{2} \},&
         &\{ \tfrac{1}{2}, 1, 1, 1 \},&
         &\{ \tfrac{1}{2}, 1, 1, \tfrac{3}{2} \},&
         &\{ \tfrac{1}{2}, 1, -1, \tfrac{3}{2} \},\\
         &\{ \tfrac{1}{2}, 1, \tfrac{3}{2}, \tfrac{3}{2} \},&
         &\{ \tfrac{2}{3}, \tfrac{2}{3}, \tfrac{2}{3}, \tfrac{2}{3} \},&
         &\{ \tfrac{2}{3}, \tfrac{2}{3}, \tfrac{2}{3}, \tfrac{4}{3} \},&
         &\{ \tfrac{2}{3}, \tfrac{2}{3}, -\tfrac{2}{3}, \tfrac{4}{3} \},&
         &\{ \tfrac{2}{3}, \tfrac{2}{3}, \tfrac{4}{3}, \tfrac{4}{3} \},\\
         &\{ \tfrac{2}{3}, -\tfrac{2}{3}, -\tfrac{2}{3}, \tfrac{4}{3} \},&
         &\{ \tfrac{2}{3}, -\tfrac{2}{3}, \tfrac{4}{3}, \tfrac{4}{3} \},&
         &\{ \tfrac{2}{3}, -\tfrac{2}{3}, \tfrac{4}{3}, \tfrac{8}{3} \},&
         &\{ \tfrac{2}{3}, -\tfrac{2}{3}, \tfrac{4}{3}, \tfrac{10}{3} \},&
         &\{ \tfrac{2}{3}, \tfrac{4}{3}, \tfrac{4}{3}, \tfrac{4}{3} \},\\
         &\{ 1, 1, 1, 1 \}; \refstepcounter{equation} \tag{\theequation}
\end{align*}
$38$ of them have $N_2 > N_0$ and no SUSY solution:
\begin{align*}
\hat r = \mbox{}
         &\{ 2, 2, 2, 0 \},&
         &\{ 2, 2, 2, -2 \},&
         &\{ 2, 2, 0, -2 \},&
         &\{ 2, 2, 0, 1 \},&
         &\{ 2, 2, -2, -2 \},\\
         &\{ 2, 2, -2, 4 \},&
         &\{ 2, 2, -2, 6 \},&
         &\{ 2, 2, \tfrac{2}{3}, -\tfrac{2}{3} \},&
         &\{ 2, 2, 1, -1 \},&
         &\{ 2, -2, -2, -2 \},\\
         &\{ 2, -2, -2, 4 \},&
         &\{ 2, -2, -2, 6 \},&
         &\{ 2, -2, \tfrac{2}{3}, -\tfrac{2}{3} \},&
         &\{ 2, -2, \tfrac{2}{3}, \tfrac{10}{3} \},&
         &\{ 2, -2, 1, -1 \},\\
         &\{ 2, -2, 1, 3 \},&
         &\{ 2, -2, -1, 4 \},&
         &\{ 2, -2, 4, 4 \},&
         &\{ 2, -2, 4, -4 \},&
         &\{ 2, -2, 4, 6 \},\\
         &\{ 2, -2, 4, -6 \},&
         &\{ 2, -2, -4, 6 \},&
         &\{ 2, -2, 6, 6 \},&
         &\{ 2, -2, 6, -10 \},&
         &\{ 2, \tfrac{2}{5}, -\tfrac{2}{5}, \tfrac{6}{5} \},\\
         &\{ 2, \tfrac{1}{2}, -\tfrac{1}{2}, 1 \},&
         &\{ 2, \tfrac{1}{2}, 1, -1 \},&
         &\{ 2, \tfrac{2}{3}, \tfrac{2}{3}, -\tfrac{2}{3} \},&
         &\{ 2, \tfrac{2}{3}, -\tfrac{2}{3}, -\tfrac{2}{3} \},&
         &\{ 2, \tfrac{2}{3}, -\tfrac{2}{3}, \tfrac{4}{3} \},\\
         &\{ 2, \tfrac{2}{3}, -\tfrac{2}{3}, \tfrac{8}{3} \},&
         &\{ 2, \tfrac{2}{3}, -\tfrac{2}{3}, \tfrac{10}{3} \},&
         &\{ 2, \tfrac{2}{3}, \tfrac{4}{3}, -\tfrac{4}{3} \},&
         &\{ 2, 1, 1, -1 \},&
         &\{ 2, 1, -1, -1 \},\\
         &\{ 2, 1, -1, \tfrac{3}{2} \},&
         &\{ 2, 1, -1, 3 \},&
         &\{ 2, 1, -1, 4 \}; \refstepcounter{equation} \tag{\theequation}
\end{align*}
and $1$ of them has $N_2 > N_0$ and a SUSY solution:
\begin{equation}
\hat r = \{ 2, -2, 6, -6 \}.
\end{equation}

\subsection{Models with $N = 5$}

There are $703$ models in total, $448$ of them have $N_2 \le N_0$ and SUSY solutions:
\begin{align*}
\hat r = \mbox{}
         &\{ 2, 2, 0, 0, 0 \},&
         &\{ 2, 2, 0, 0, -2 \},&
         &\{ 2, 2, 0, 0, 1 \},&
         &\{ 2, 0, 0, 0, 0 \},\\
         &\{ 2, 0, 0, 0, -2 \},&
         &\{ 2, 0, 0, 0, 1 \},&
         &\{ 2, 0, 0, -2, -2 \},&
         &\{ 2, 0, 0, -2, 1 \},\\
         &\{ 2, 0, 0, -2, 4 \},&
         &\{ 2, 0, 0, -2, 6 \},&
         &\{ 2, 0, 0, \tfrac{1}{2}, 1 \},&
         &\{ 2, 0, 0, \tfrac{2}{3}, -\tfrac{2}{3} \},\\
         &\{ 2, 0, 0, \tfrac{2}{3}, \tfrac{4}{3} \},&
         &\{ 2, 0, 0, 1, 1 \},&
         &\{ 2, 0, 0, 1, -1 \},&
         &\{ 2, 0, -2, -2, -2 \},\\
         &\{ 2, 0, -2, -2, 1 \},&
         &\{ 2, 0, -2, -2, 4 \},&
         &\{ 2, 0, -2, -2, 6 \},&
         &\{ 2, 0, -2, \tfrac{1}{2}, 1 \},\\
         &\{ 2, 0, -2, \tfrac{2}{3}, -\tfrac{2}{3} \},&
         &\{ 2, 0, -2, \tfrac{2}{3}, \tfrac{4}{3} \},&
         &\{ 2, 0, -2, \tfrac{2}{3}, \tfrac{10}{3} \},&
         &\{ 2, 0, -2, 1, 1 \},\\
         &\{ 2, 0, -2, 1, -1 \},&
         &\{ 2, 0, -2, 1, 3 \},&
         &\{ 2, 0, -2, 1, 4 \},&
         &\{ 2, 0, -2, 1, 6 \},\\
         &\{ 2, 0, -2, -1, 4 \},&
         &\{ 2, 0, -2, 4, 4 \},&
         &\{ 2, 0, -2, 4, -4 \},&
         &\{ 2, 0, -2, 4, 6 \},\\
         &\{ 2, 0, -2, 4, -6 \},&
         &\{ 2, 0, -2, -4, 6 \},&
         &\{ 2, 0, -2, 6, 6 \},&
         &\{ 2, 0, -2, 6, -6 \},\\
         &\{ 2, 0, -2, 6, -10 \},&
         &\{ 2, 0, \tfrac{1}{3}, \tfrac{2}{3}, 1 \},&
         &\{ 2, 0, \tfrac{1}{3}, \tfrac{2}{3}, \tfrac{4}{3} \},&
         &\{ 2, 0, \tfrac{2}{5}, -\tfrac{2}{5}, \tfrac{6}{5} \},\\
         &\{ 2, 0, \tfrac{2}{5}, \tfrac{4}{5}, \tfrac{6}{5} \},&
         &\{ 2, 0, \tfrac{1}{2}, \tfrac{1}{2}, 1 \},&
         &\{ 2, 0, \tfrac{1}{2}, -\tfrac{1}{2}, 1 \},&
         &\{ 2, 0, \tfrac{1}{2}, \tfrac{3}{4}, 1 \},\\
         &\{ 2, 0, \tfrac{1}{2}, 1, 1 \},&
         &\{ 2, 0, \tfrac{1}{2}, 1, -1 \},&
         &\{ 2, 0, \tfrac{1}{2}, 1, \tfrac{3}{2} \},&
         &\{ 2, 0, \tfrac{2}{3}, \tfrac{2}{3}, -\tfrac{2}{3} \},\\
         &\{ 2, 0, \tfrac{2}{3}, \tfrac{2}{3}, \tfrac{4}{3} \},&
         &\{ 2, 0, \tfrac{2}{3}, -\tfrac{2}{3}, -\tfrac{2}{3} \},&
         &\{ 2, 0, \tfrac{2}{3}, -\tfrac{2}{3}, 1 \},&
         &\{ 2, 0, \tfrac{2}{3}, -\tfrac{2}{3}, \tfrac{4}{3} \},\\
         &\{ 2, 0, \tfrac{2}{3}, -\tfrac{2}{3}, \tfrac{8}{3} \},&
         &\{ 2, 0, \tfrac{2}{3}, -\tfrac{2}{3}, \tfrac{10}{3} \},&
         &\{ 2, 0, \tfrac{2}{3}, 1, \tfrac{4}{3} \},&
         &\{ 2, 0, \tfrac{2}{3}, \tfrac{4}{3}, \tfrac{4}{3} \},\\
         &\{ 2, 0, \tfrac{2}{3}, \tfrac{4}{3}, -\tfrac{4}{3} \},&
         &\{ 2, 0, 1, 1, 1 \},&
         &\{ 2, 0, 1, 1, -1 \},&
         &\{ 2, 0, 1, -1, -1 \},\\
         &\{ 2, 0, 1, -1, \tfrac{3}{2} \},&
         &\{ 2, 0, 1, -1, 3 \},&
         &\{ 2, 0, 1, -1, 4 \},&
         &\{ 0, 0, 0, 0, 1 \},\\
         &\{ 0, 0, 0, \tfrac{1}{2}, 1 \},&
         &\{ 0, 0, 0, \tfrac{2}{3}, \tfrac{4}{3} \},&
         &\{ 0, 0, 0, 1, 1 \},&
         &\{ 0, 0, \tfrac{1}{3}, \tfrac{2}{3}, 1 \},\\
         &\{ 0, 0, \tfrac{1}{3}, \tfrac{2}{3}, \tfrac{4}{3} \},&
         &\{ 0, 0, \tfrac{2}{5}, \tfrac{4}{5}, \tfrac{6}{5} \},&
         &\{ 0, 0, \tfrac{1}{2}, \tfrac{1}{2}, 1 \},&
         &\{ 0, 0, \tfrac{1}{2}, \tfrac{3}{4}, 1 \},\\
         &\{ 0, 0, \tfrac{1}{2}, 1, 1 \},&
         &\{ 0, 0, \tfrac{1}{2}, 1, \tfrac{3}{2} \},&
         &\{ 0, 0, \tfrac{2}{3}, \tfrac{2}{3}, \tfrac{4}{3} \},&
         &\{ 0, 0, \tfrac{2}{3}, -\tfrac{2}{3}, \tfrac{4}{3} \},\\
         &\{ 0, 0, \tfrac{2}{3}, 1, \tfrac{4}{3} \},&
         &\{ 0, 0, \tfrac{2}{3}, \tfrac{4}{3}, \tfrac{4}{3} \},&
         &\{ 0, 0, 1, 1, 1 \},&
         &\{ 0, \tfrac{2}{9}, \tfrac{2}{3}, \tfrac{8}{9}, \tfrac{10}{9} \},\\
         &\{ 0, \tfrac{1}{4}, \tfrac{1}{2}, \tfrac{3}{4}, 1 \},&
         &\{ 0, \tfrac{1}{4}, \tfrac{1}{2}, 1, \tfrac{3}{2} \},&
         &\{ 0, \tfrac{2}{7}, \tfrac{4}{7}, \tfrac{6}{7}, \tfrac{8}{7} \},&
         &\{ 0, \tfrac{2}{7}, \tfrac{4}{7}, \tfrac{6}{7}, \tfrac{10}{7} \},\\
         &\{ 0, \tfrac{1}{3}, \tfrac{1}{3}, \tfrac{2}{3}, 1 \},&
         &\{ 0, \tfrac{1}{3}, \tfrac{1}{3}, \tfrac{2}{3}, \tfrac{4}{3} \},&
         &\{ 0, \tfrac{1}{3}, \tfrac{1}{2}, \tfrac{2}{3}, 1 \},&
         &\{ 0, \tfrac{1}{3}, \tfrac{2}{3}, \tfrac{2}{3}, 1 \},\\
         &\{ 0, \tfrac{1}{3}, \tfrac{2}{3}, \tfrac{2}{3}, \tfrac{4}{3} \},&
         &\{ 0, \tfrac{1}{3}, \tfrac{2}{3}, -\tfrac{2}{3}, \tfrac{4}{3} \},&
         &\{ 0, \tfrac{1}{3}, \tfrac{2}{3}, \tfrac{5}{6}, 1 \},&
         &\{ 0, \tfrac{1}{3}, \tfrac{2}{3}, \tfrac{5}{6}, \tfrac{4}{3} \},\\
         &\{ 0, \tfrac{1}{3}, \tfrac{2}{3}, 1, 1 \},&
         &\{ 0, \tfrac{1}{3}, \tfrac{2}{3}, 1, \tfrac{4}{3} \},&
         &\{ 0, \tfrac{1}{3}, \tfrac{2}{3}, 1, \tfrac{5}{3} \},&
         &\{ 0, \tfrac{1}{3}, \tfrac{2}{3}, \tfrac{4}{3}, \tfrac{4}{3} \},\\
         &\{ 0, \tfrac{1}{3}, \tfrac{2}{3}, \tfrac{4}{3}, \tfrac{5}{3} \},&
         &\{ 0, -\tfrac{1}{3}, \tfrac{2}{3}, 1, \tfrac{4}{3} \},&
         &\{ 0, \tfrac{2}{5}, \tfrac{2}{5}, \tfrac{4}{5}, \tfrac{6}{5} \},&
         &\{ 0, \tfrac{2}{5}, -\tfrac{2}{5}, \tfrac{4}{5}, \tfrac{6}{5} \},\\
         &\{ 0, \tfrac{2}{5}, \tfrac{3}{5}, \tfrac{4}{5}, 1 \},&
         &\{ 0, \tfrac{2}{5}, \tfrac{3}{5}, \tfrac{4}{5}, \tfrac{6}{5} \},&
         &\{ 0, \tfrac{2}{5}, \tfrac{4}{5}, \tfrac{4}{5}, \tfrac{6}{5} \},&
         &\{ 0, \tfrac{2}{5}, \tfrac{4}{5}, 1, \tfrac{6}{5} \},\\
         &\{ 0, \tfrac{2}{5}, \tfrac{4}{5}, \tfrac{6}{5}, \tfrac{6}{5} \},&
         &\{ 0, \tfrac{2}{5}, \tfrac{4}{5}, \tfrac{6}{5}, \tfrac{8}{5} \},&
         &\{ 0, \tfrac{4}{9}, \tfrac{2}{3}, \tfrac{8}{9}, \tfrac{10}{9} \},&
         &\{ 0, \tfrac{1}{2}, \tfrac{1}{2}, \tfrac{1}{2}, 1 \},\\
         &\{ 0, \tfrac{1}{2}, \tfrac{1}{2}, \tfrac{3}{4}, 1 \},&
         &\{ 0, \tfrac{1}{2}, \tfrac{1}{2}, 1, 1 \},&
         &\{ 0, \tfrac{1}{2}, \tfrac{1}{2}, 1, \tfrac{3}{2} \},&
         &\{ 0, \tfrac{1}{2}, -\tfrac{1}{2}, 1, \tfrac{3}{2} \},\\
         &\{ 0, \tfrac{1}{2}, \tfrac{5}{8}, \tfrac{3}{4}, 1 \},&
         &\{ 0, \tfrac{1}{2}, \tfrac{2}{3}, \tfrac{5}{6}, 1 \},&
         &\{ 0, \tfrac{1}{2}, \tfrac{2}{3}, 1, \tfrac{4}{3} \},&
         &\{ 0, \tfrac{1}{2}, \tfrac{3}{4}, \tfrac{3}{4}, 1 \},\\
         &\{ 0, \tfrac{1}{2}, \tfrac{3}{4}, 1, 1 \},&
         &\{ 0, \tfrac{1}{2}, \tfrac{3}{4}, 1, \tfrac{5}{4} \},&
         &\{ 0, \tfrac{1}{2}, \tfrac{3}{4}, 1, \tfrac{3}{2} \},&
         &\{ 0, \tfrac{1}{2}, 1, 1, 1 \},\\
         &\{ 0, \tfrac{1}{2}, 1, 1, \tfrac{3}{2} \},&
         &\{ 0, \tfrac{1}{2}, 1, -1, \tfrac{3}{2} \},&
         &\{ 0, \tfrac{1}{2}, 1, \tfrac{3}{2}, \tfrac{3}{2} \},&
         &\{ 0, \tfrac{2}{3}, \tfrac{2}{3}, \tfrac{2}{3}, \tfrac{4}{3} \},\\
         &\{ 0, \tfrac{2}{3}, \tfrac{2}{3}, -\tfrac{2}{3}, \tfrac{4}{3} \},&
         &\{ 0, \tfrac{2}{3}, \tfrac{2}{3}, 1, \tfrac{4}{3} \},&
         &\{ 0, \tfrac{2}{3}, \tfrac{2}{3}, \tfrac{4}{3}, \tfrac{4}{3} \},&
         &\{ 0, \tfrac{2}{3}, -\tfrac{2}{3}, -\tfrac{2}{3}, \tfrac{4}{3} \},\\
         &\{ 0, \tfrac{2}{3}, -\tfrac{2}{3}, 1, \tfrac{4}{3} \},&
         &\{ 0, \tfrac{2}{3}, -\tfrac{2}{3}, \tfrac{4}{3}, \tfrac{4}{3} \},&
         &\{ 0, \tfrac{2}{3}, -\tfrac{2}{3}, \tfrac{4}{3}, \tfrac{8}{3} \},&
         &\{ 0, \tfrac{2}{3}, -\tfrac{2}{3}, \tfrac{4}{3}, \tfrac{10}{3} \},\\
         &\{ 0, \tfrac{2}{3}, 1, 1, \tfrac{4}{3} \},&
         &\{ 0, \tfrac{2}{3}, 1, \tfrac{4}{3}, \tfrac{4}{3} \},&
         &\{ 0, \tfrac{2}{3}, \tfrac{4}{3}, \tfrac{4}{3}, \tfrac{4}{3} \},&
         &\{ 0, 1, 1, 1, 1 \},\\
         &\{ -2, \tfrac{2}{3}, -\tfrac{2}{3}, \tfrac{4}{3}, \tfrac{8}{3} \},&
         &\{ -2, \tfrac{2}{3}, -\tfrac{2}{3}, \tfrac{4}{3}, \tfrac{10}{3} \},&
         &\{ -2, 1, -1, 3, 4 \},&
         &\{ \tfrac{2}{15}, \tfrac{2}{5}, \tfrac{2}{3}, \tfrac{4}{5}, \tfrac{6}{5} \},\\
         &\{ \tfrac{2}{15}, \tfrac{8}{15}, \tfrac{2}{3}, \tfrac{14}{15}, \tfrac{4}{3} \},&
         &\{ \tfrac{2}{13}, \tfrac{6}{13}, \tfrac{10}{13}, \tfrac{12}{13}, \tfrac{14}{13} \},&
         &\{ \tfrac{1}{6}, \tfrac{1}{3}, \tfrac{2}{3}, \tfrac{5}{6}, 1 \},&
         &\{ \tfrac{1}{6}, \tfrac{1}{3}, \tfrac{2}{3}, 1, \tfrac{5}{3} \},\\
         &\{ \tfrac{1}{6}, \tfrac{1}{3}, \tfrac{2}{3}, \tfrac{4}{3}, \tfrac{5}{3} \},&
         &\{ \tfrac{1}{6}, \tfrac{1}{2}, \tfrac{2}{3}, \tfrac{5}{6}, 1 \},&
         &\{ \tfrac{1}{6}, \tfrac{1}{2}, \tfrac{2}{3}, 1, \tfrac{4}{3} \},&
         &\{ \tfrac{1}{6}, \tfrac{2}{3}, \tfrac{5}{6}, 1, \tfrac{7}{6} \},\\
         &\{ -\tfrac{1}{6}, \tfrac{1}{3}, \tfrac{2}{3}, \tfrac{5}{6}, \tfrac{4}{3} \},&
         &\{ -\tfrac{1}{6}, \tfrac{1}{2}, \tfrac{2}{3}, 1, \tfrac{3}{2} \},&
         &\{ \tfrac{2}{11}, \tfrac{6}{11}, \tfrac{8}{11}, \tfrac{10}{11}, \tfrac{12}{11} \},&
         &\{ \tfrac{2}{11}, \tfrac{6}{11}, \tfrac{8}{11}, \tfrac{10}{11}, \tfrac{14}{11} \},\\
         &\{ \tfrac{1}{5}, \tfrac{2}{5}, \tfrac{3}{5}, \tfrac{4}{5}, 1 \},&
         &\{ \tfrac{1}{5}, \tfrac{2}{5}, \tfrac{3}{5}, \tfrac{4}{5}, \tfrac{6}{5} \},&
         &\{ \tfrac{1}{5}, \tfrac{2}{5}, \tfrac{4}{5}, 1, \tfrac{6}{5} \},&
         &\{ \tfrac{1}{5}, \tfrac{2}{5}, \tfrac{4}{5}, \tfrac{6}{5}, \tfrac{8}{5} \},\\
         &\{ \tfrac{1}{5}, \tfrac{3}{5}, \tfrac{4}{5}, 1, \tfrac{6}{5} \},&
         &\{ -\tfrac{1}{5}, \tfrac{2}{5}, \tfrac{4}{5}, 1, \tfrac{6}{5} \},&
         &\{ \tfrac{2}{9}, \tfrac{2}{9}, \tfrac{2}{3}, \tfrac{8}{9}, \tfrac{10}{9} \},&
         &\{ \tfrac{2}{9}, -\tfrac{2}{9}, \tfrac{2}{3}, \tfrac{8}{9}, \tfrac{10}{9} \},\\
         &\{ \tfrac{2}{9}, \tfrac{4}{9}, \tfrac{2}{3}, \tfrac{8}{9}, \tfrac{10}{9} \},&
         &\{ \tfrac{2}{9}, \tfrac{4}{9}, \tfrac{2}{3}, \tfrac{8}{9}, \tfrac{4}{3} \},&
         &\{ \tfrac{2}{9}, \tfrac{4}{9}, \tfrac{2}{3}, \tfrac{8}{9}, \tfrac{14}{9} \},&
         &\{ \tfrac{2}{9}, \tfrac{4}{9}, \tfrac{2}{3}, \tfrac{10}{9}, \tfrac{4}{3} \},\\
         &\{ \tfrac{2}{9}, \tfrac{4}{9}, \tfrac{2}{3}, \tfrac{10}{9}, \tfrac{14}{9} \},&
         &\{ \tfrac{2}{9}, \tfrac{4}{9}, \tfrac{2}{3}, \tfrac{4}{3}, \tfrac{14}{9} \},&
         &\{ \tfrac{2}{9}, \tfrac{4}{9}, \tfrac{8}{9}, \tfrac{10}{9}, \tfrac{14}{9} \},&
         &\{ \tfrac{2}{9}, \tfrac{5}{9}, \tfrac{2}{3}, \tfrac{8}{9}, \tfrac{10}{9} \},\\
         &\{ \tfrac{2}{9}, \tfrac{2}{3}, \tfrac{2}{3}, \tfrac{8}{9}, \tfrac{10}{9} \},&
         &\{ \tfrac{2}{9}, \tfrac{2}{3}, \tfrac{8}{9}, \tfrac{8}{9}, \tfrac{10}{9} \},&
         &\{ \tfrac{2}{9}, \tfrac{2}{3}, \tfrac{8}{9}, \tfrac{10}{9}, \tfrac{10}{9} \},&
         &\{ \tfrac{2}{9}, \tfrac{2}{3}, \tfrac{8}{9}, \tfrac{10}{9}, \tfrac{4}{3} \},\\
         &\{ \tfrac{2}{9}, \tfrac{2}{3}, \tfrac{8}{9}, \tfrac{10}{9}, \tfrac{14}{9} \},&
         &\{ \tfrac{2}{9}, \tfrac{2}{3}, \tfrac{8}{9}, \tfrac{10}{9}, \tfrac{16}{9} \},&
         &\{ -\tfrac{2}{9}, \tfrac{4}{9}, \tfrac{2}{3}, \tfrac{8}{9}, \tfrac{10}{9} \},&
         &\{ -\tfrac{2}{9}, \tfrac{4}{9}, \tfrac{2}{3}, \tfrac{10}{9}, \tfrac{14}{9} \},\\
         &\{ -\tfrac{2}{9}, \tfrac{2}{3}, \tfrac{8}{9}, \tfrac{10}{9}, \tfrac{4}{3} \},&
         &\{ \tfrac{1}{4}, \tfrac{1}{4}, \tfrac{1}{2}, \tfrac{3}{4}, 1 \},&
         &\{ \tfrac{1}{4}, \tfrac{1}{4}, \tfrac{1}{2}, 1, \tfrac{3}{2} \},&
         &\{ \tfrac{1}{4}, \tfrac{1}{2}, \tfrac{1}{2}, \tfrac{3}{4}, 1 \},\\
         &\{ \tfrac{1}{4}, \tfrac{1}{2}, \tfrac{1}{2}, 1, \tfrac{3}{2} \},&
         &\{ \tfrac{1}{4}, \tfrac{1}{2}, -\tfrac{1}{2}, 1, \tfrac{3}{2} \},&
         &\{ \tfrac{1}{4}, \tfrac{1}{2}, \tfrac{5}{8}, \tfrac{3}{4}, 1 \},&
         &\{ \tfrac{1}{4}, \tfrac{1}{2}, \tfrac{3}{4}, \tfrac{3}{4}, 1 \},\\
         &\{ \tfrac{1}{4}, \tfrac{1}{2}, \tfrac{3}{4}, \tfrac{7}{8}, 1 \},&
         &\{ \tfrac{1}{4}, \tfrac{1}{2}, \tfrac{3}{4}, 1, 1 \},&
         &\{ \tfrac{1}{4}, \tfrac{1}{2}, \tfrac{3}{4}, 1, \tfrac{5}{4} \},&
         &\{ \tfrac{1}{4}, \tfrac{1}{2}, \tfrac{3}{4}, 1, \tfrac{3}{2} \},\\
         &\{ \tfrac{1}{4}, \tfrac{1}{2}, \tfrac{3}{4}, 1, \tfrac{7}{4} \},&
         &\{ \tfrac{1}{4}, \tfrac{1}{2}, \tfrac{3}{4}, \tfrac{5}{4}, \tfrac{3}{2} \},&
         &\{ \tfrac{1}{4}, \tfrac{1}{2}, \tfrac{7}{8}, 1, \tfrac{3}{2} \},&
         &\{ \tfrac{1}{4}, \tfrac{1}{2}, 1, 1, \tfrac{3}{2} \},\\
         &\{ \tfrac{1}{4}, \tfrac{1}{2}, 1, -1, \tfrac{3}{2} \},&
         &\{ \tfrac{1}{4}, \tfrac{1}{2}, 1, \tfrac{5}{4}, \tfrac{3}{2} \},&
         &\{ \tfrac{1}{4}, \tfrac{1}{2}, 1, \tfrac{3}{2}, \tfrac{3}{2} \},&
         &\{ \tfrac{1}{4}, \tfrac{1}{2}, 1, \tfrac{3}{2}, \tfrac{7}{4} \},\\
         &\{ -\tfrac{1}{4}, \tfrac{1}{2}, \tfrac{3}{4}, 1, \tfrac{5}{4} \},&
         &\{ -\tfrac{1}{4}, \tfrac{1}{2}, \tfrac{3}{4}, 1, \tfrac{3}{2} \},&
         &\{ \tfrac{2}{7}, \tfrac{2}{7}, \tfrac{4}{7}, \tfrac{6}{7}, \tfrac{8}{7} \},&
         &\{ \tfrac{2}{7}, \tfrac{2}{7}, \tfrac{4}{7}, \tfrac{6}{7}, \tfrac{10}{7} \},\\
         &\{ \tfrac{2}{7}, -\tfrac{2}{7}, \tfrac{4}{7}, \tfrac{6}{7}, \tfrac{8}{7} \},&
         &\{ \tfrac{2}{7}, -\tfrac{2}{7}, \tfrac{4}{7}, \tfrac{6}{7}, \tfrac{10}{7} \},&
         &\{ \tfrac{2}{7}, -\tfrac{2}{7}, \tfrac{6}{7}, \tfrac{8}{7}, \tfrac{10}{7} \},&
         &\{ \tfrac{2}{7}, \tfrac{3}{7}, \tfrac{4}{7}, \tfrac{6}{7}, \tfrac{8}{7} \},\\
         &\{ \tfrac{2}{7}, \tfrac{4}{7}, \tfrac{4}{7}, \tfrac{6}{7}, \tfrac{8}{7} \},&
         &\{ \tfrac{2}{7}, \tfrac{4}{7}, \tfrac{4}{7}, \tfrac{6}{7}, \tfrac{10}{7} \},&
         &\{ \tfrac{2}{7}, \tfrac{4}{7}, \tfrac{5}{7}, \tfrac{6}{7}, 1 \},&
         &\{ \tfrac{2}{7}, \tfrac{4}{7}, \tfrac{5}{7}, \tfrac{6}{7}, \tfrac{8}{7} \},\\
         &\{ \tfrac{2}{7}, \tfrac{4}{7}, \tfrac{5}{7}, \tfrac{6}{7}, \tfrac{10}{7} \},&
         &\{ \tfrac{2}{7}, \tfrac{4}{7}, \tfrac{6}{7}, \tfrac{6}{7}, \tfrac{8}{7} \},&
         &\{ \tfrac{2}{7}, \tfrac{4}{7}, \tfrac{6}{7}, \tfrac{6}{7}, \tfrac{10}{7} \},&
         &\{ \tfrac{2}{7}, \tfrac{4}{7}, \tfrac{6}{7}, -\tfrac{6}{7}, \tfrac{10}{7} \},\\
         &\{ \tfrac{2}{7}, \tfrac{4}{7}, \tfrac{6}{7}, \tfrac{8}{7}, \tfrac{8}{7} \},&
         &\{ \tfrac{2}{7}, \tfrac{4}{7}, \tfrac{6}{7}, \tfrac{8}{7}, \tfrac{10}{7} \},&
         &\{ \tfrac{2}{7}, \tfrac{4}{7}, \tfrac{6}{7}, \tfrac{8}{7}, \tfrac{12}{7} \},&
         &\{ \tfrac{2}{7}, \tfrac{4}{7}, \tfrac{6}{7}, \tfrac{10}{7}, \tfrac{10}{7} \},\\
         &\{ \tfrac{2}{7}, \tfrac{4}{7}, \tfrac{6}{7}, \tfrac{10}{7}, \tfrac{12}{7} \},&
         &\{ -\tfrac{2}{7}, \tfrac{4}{7}, \tfrac{6}{7}, \tfrac{8}{7}, \tfrac{10}{7} \},&
         &\{ \tfrac{1}{3}, \tfrac{1}{3}, \tfrac{1}{3}, \tfrac{2}{3}, 1 \},&
         &\{ \tfrac{1}{3}, \tfrac{1}{3}, \tfrac{1}{3}, \tfrac{2}{3}, \tfrac{4}{3} \},\\
         &\{ \tfrac{1}{3}, \tfrac{1}{3}, \tfrac{1}{2}, \tfrac{2}{3}, 1 \},&
         &\{ \tfrac{1}{3}, \tfrac{1}{3}, \tfrac{2}{3}, \tfrac{2}{3}, 1 \},&
         &\{ \tfrac{1}{3}, \tfrac{1}{3}, \tfrac{2}{3}, \tfrac{2}{3}, \tfrac{4}{3} \},&
         &\{ \tfrac{1}{3}, \tfrac{1}{3}, \tfrac{2}{3}, -\tfrac{2}{3}, \tfrac{4}{3} \},\\
         &\{ \tfrac{1}{3}, \tfrac{1}{3}, \tfrac{2}{3}, \tfrac{5}{6}, 1 \},&
         &\{ \tfrac{1}{3}, \tfrac{1}{3}, \tfrac{2}{3}, \tfrac{5}{6}, \tfrac{4}{3} \},&
         &\{ \tfrac{1}{3}, \tfrac{1}{3}, \tfrac{2}{3}, 1, 1 \},&
         &\{ \tfrac{1}{3}, \tfrac{1}{3}, \tfrac{2}{3}, 1, \tfrac{4}{3} \},\\
         &\{ \tfrac{1}{3}, \tfrac{1}{3}, \tfrac{2}{3}, 1, \tfrac{5}{3} \},&
         &\{ \tfrac{1}{3}, \tfrac{1}{3}, \tfrac{2}{3}, \tfrac{4}{3}, \tfrac{4}{3} \},&
         &\{ \tfrac{1}{3}, \tfrac{1}{3}, \tfrac{2}{3}, \tfrac{4}{3}, \tfrac{5}{3} \},&
         &\{ \tfrac{1}{3}, -\tfrac{1}{3}, \tfrac{2}{3}, 1, \tfrac{4}{3} \},\\
         &\{ \tfrac{1}{3}, -\tfrac{1}{3}, \tfrac{2}{3}, 1, \tfrac{5}{3} \},&
         &\{ \tfrac{1}{3}, -\tfrac{1}{3}, \tfrac{2}{3}, \tfrac{4}{3}, \tfrac{5}{3} \},&
         &\{ \tfrac{1}{3}, \tfrac{1}{2}, \tfrac{1}{2}, \tfrac{2}{3}, 1 \},&
         &\{ \tfrac{1}{3}, \tfrac{1}{2}, \tfrac{2}{3}, \tfrac{2}{3}, 1 \},\\
         &\{ \tfrac{1}{3}, \tfrac{1}{2}, \tfrac{2}{3}, \tfrac{3}{4}, 1 \},&
         &\{ \tfrac{1}{3}, \tfrac{1}{2}, \tfrac{2}{3}, \tfrac{5}{6}, 1 \},&
         &\{ \tfrac{1}{3}, \tfrac{1}{2}, \tfrac{2}{3}, \tfrac{5}{6}, \tfrac{7}{6} \},&
         &\{ \tfrac{1}{3}, \tfrac{1}{2}, \tfrac{2}{3}, \tfrac{5}{6}, \tfrac{4}{3} \},\\
         &\{ \tfrac{1}{3}, \tfrac{1}{2}, \tfrac{2}{3}, 1, 1 \},&
         &\{ \tfrac{1}{3}, \tfrac{1}{2}, \tfrac{2}{3}, 1, \tfrac{7}{6} \},&
         &\{ \tfrac{1}{3}, \tfrac{1}{2}, \tfrac{2}{3}, 1, \tfrac{4}{3} \},&
         &\{ \tfrac{1}{3}, \tfrac{1}{2}, \tfrac{2}{3}, 1, \tfrac{3}{2} \},\\
         &\{ \tfrac{1}{3}, \tfrac{1}{2}, \tfrac{2}{3}, 1, \tfrac{5}{3} \},&
         &\{ \tfrac{1}{3}, \tfrac{1}{2}, \tfrac{5}{6}, 1, \tfrac{7}{6} \},&
         &\{ \tfrac{1}{3}, \tfrac{7}{12}, \tfrac{2}{3}, \tfrac{5}{6}, 1 \},&
         &\{ \tfrac{1}{3}, \tfrac{7}{12}, \tfrac{2}{3}, \tfrac{5}{6}, \tfrac{4}{3} \},\\
         &\{ \tfrac{1}{3}, \tfrac{2}{3}, \tfrac{2}{3}, \tfrac{2}{3}, 1 \},&
         &\{ \tfrac{1}{3}, \tfrac{2}{3}, \tfrac{2}{3}, \tfrac{2}{3}, \tfrac{4}{3} \},&
         &\{ \tfrac{1}{3}, \tfrac{2}{3}, \tfrac{2}{3}, -\tfrac{2}{3}, \tfrac{4}{3} \},&
         &\{ \tfrac{1}{3}, \tfrac{2}{3}, \tfrac{2}{3}, \tfrac{5}{6}, 1 \},\\
         &\{ \tfrac{1}{3}, \tfrac{2}{3}, \tfrac{2}{3}, \tfrac{5}{6}, \tfrac{4}{3} \},&
         &\{ \tfrac{1}{3}, \tfrac{2}{3}, \tfrac{2}{3}, 1, 1 \},&
         &\{ \tfrac{1}{3}, \tfrac{2}{3}, \tfrac{2}{3}, 1, \tfrac{4}{3} \},&
         &\{ \tfrac{1}{3}, \tfrac{2}{3}, \tfrac{2}{3}, 1, \tfrac{5}{3} \},\\
         &\{ \tfrac{1}{3}, \tfrac{2}{3}, \tfrac{2}{3}, \tfrac{4}{3}, \tfrac{4}{3} \},&
         &\{ \tfrac{1}{3}, \tfrac{2}{3}, \tfrac{2}{3}, \tfrac{4}{3}, \tfrac{5}{3} \},&
         &\{ \tfrac{1}{3}, \tfrac{2}{3}, -\tfrac{2}{3}, -\tfrac{2}{3}, \tfrac{4}{3} \},&
         &\{ \tfrac{1}{3}, \tfrac{2}{3}, -\tfrac{2}{3}, \tfrac{5}{6}, \tfrac{4}{3} \},\\
         &\{ \tfrac{1}{3}, \tfrac{2}{3}, -\tfrac{2}{3}, 1, \tfrac{4}{3} \},&
         &\{ \tfrac{1}{3}, \tfrac{2}{3}, -\tfrac{2}{3}, 1, \tfrac{5}{3} \},&
         &\{ \tfrac{1}{3}, \tfrac{2}{3}, -\tfrac{2}{3}, \tfrac{4}{3}, \tfrac{4}{3} \},&
         &\{ \tfrac{1}{3}, \tfrac{2}{3}, -\tfrac{2}{3}, \tfrac{4}{3}, \tfrac{5}{3} \},\\
         &\{ \tfrac{1}{3}, \tfrac{2}{3}, -\tfrac{2}{3}, \tfrac{4}{3}, \tfrac{7}{3} \},&
         &\{ \tfrac{1}{3}, \tfrac{2}{3}, -\tfrac{2}{3}, \tfrac{4}{3}, \tfrac{8}{3} \},&
         &\{ \tfrac{1}{3}, \tfrac{2}{3}, -\tfrac{2}{3}, \tfrac{4}{3}, \tfrac{10}{3} \},&
         &\{ \tfrac{1}{3}, \tfrac{2}{3}, \tfrac{5}{6}, \tfrac{5}{6}, 1 \},\\
         &\{ \tfrac{1}{3}, \tfrac{2}{3}, \tfrac{5}{6}, \tfrac{5}{6}, \tfrac{4}{3} \},&
         &\{ \tfrac{1}{3}, \tfrac{2}{3}, \tfrac{5}{6}, 1, 1 \},&
         &\{ \tfrac{1}{3}, \tfrac{2}{3}, \tfrac{5}{6}, 1, \tfrac{7}{6} \},&
         &\{ \tfrac{1}{3}, \tfrac{2}{3}, \tfrac{5}{6}, 1, \tfrac{4}{3} \},\\
         &\{ \tfrac{1}{3}, \tfrac{2}{3}, \tfrac{5}{6}, 1, \tfrac{5}{3} \},&
         &\{ \tfrac{1}{3}, \tfrac{2}{3}, \tfrac{5}{6}, \tfrac{7}{6}, \tfrac{4}{3} \},&
         &\{ \tfrac{1}{3}, \tfrac{2}{3}, \tfrac{5}{6}, \tfrac{4}{3}, \tfrac{4}{3} \},&
         &\{ \tfrac{1}{3}, \tfrac{2}{3}, \tfrac{5}{6}, \tfrac{4}{3}, \tfrac{5}{3} \},\\
         &\{ \tfrac{1}{3}, \tfrac{2}{3}, 1, 1, 1 \},&
         &\{ \tfrac{1}{3}, \tfrac{2}{3}, 1, 1, \tfrac{4}{3} \},&
         &\{ \tfrac{1}{3}, \tfrac{2}{3}, 1, 1, \tfrac{5}{3} \},&
         &\{ \tfrac{1}{3}, \tfrac{2}{3}, 1, \tfrac{4}{3}, \tfrac{4}{3} \},\\
         &\{ \tfrac{1}{3}, \tfrac{2}{3}, 1, \tfrac{4}{3}, \tfrac{5}{3} \},&
         &\{ \tfrac{1}{3}, \tfrac{2}{3}, 1, -\tfrac{4}{3}, \tfrac{5}{3} \},&
         &\{ \tfrac{1}{3}, \tfrac{2}{3}, 1, \tfrac{5}{3}, \tfrac{5}{3} \},&
         &\{ \tfrac{1}{3}, \tfrac{2}{3}, -1, \tfrac{4}{3}, \tfrac{5}{3} \},\\
         &\{ \tfrac{1}{3}, \tfrac{2}{3}, \tfrac{4}{3}, \tfrac{4}{3}, \tfrac{4}{3} \},&
         &\{ \tfrac{1}{3}, \tfrac{2}{3}, \tfrac{4}{3}, \tfrac{4}{3}, \tfrac{5}{3} \},&
         &\{ \tfrac{1}{3}, \tfrac{2}{3}, \tfrac{4}{3}, -\tfrac{4}{3}, \tfrac{5}{3} \},&
         &\{ \tfrac{1}{3}, \tfrac{2}{3}, \tfrac{4}{3}, \tfrac{5}{3}, \tfrac{5}{3} \},\\
         &\{ \tfrac{1}{3}, -\tfrac{2}{3}, 1, \tfrac{4}{3}, \tfrac{5}{3} \},&
         &\{ -\tfrac{1}{3}, -\tfrac{1}{3}, \tfrac{2}{3}, 1, \tfrac{4}{3} \},&
         &\{ -\tfrac{1}{3}, \tfrac{1}{2}, \tfrac{2}{3}, 1, \tfrac{4}{3} \},&
         &\{ -\tfrac{1}{3}, \tfrac{2}{3}, \tfrac{2}{3}, 1, \tfrac{4}{3} \},\\
         &\{ -\tfrac{1}{3}, \tfrac{2}{3}, -\tfrac{2}{3}, 1, \tfrac{4}{3} \},&
         &\{ -\tfrac{1}{3}, \tfrac{2}{3}, -\tfrac{2}{3}, \tfrac{4}{3}, \tfrac{8}{3} \},&
         &\{ -\tfrac{1}{3}, \tfrac{2}{3}, 1, 1, \tfrac{4}{3} \},&
         &\{ -\tfrac{1}{3}, \tfrac{2}{3}, 1, \tfrac{7}{6}, \tfrac{4}{3} \},\\
         &\{ -\tfrac{1}{3}, \tfrac{2}{3}, 1, \tfrac{4}{3}, \tfrac{4}{3} \},&
         &\{ -\tfrac{1}{3}, \tfrac{2}{3}, 1, \tfrac{4}{3}, \tfrac{5}{3} \},&
         &\{ -\tfrac{1}{3}, \tfrac{2}{3}, 1, \tfrac{4}{3}, \tfrac{7}{3} \},&
         &\{ -\tfrac{1}{3}, \tfrac{2}{3}, 1, \tfrac{4}{3}, \tfrac{8}{3} \},\\
         &\{ \tfrac{6}{17}, \tfrac{10}{17}, \tfrac{12}{17}, \tfrac{14}{17}, \tfrac{22}{17} \},&
         &\{ \tfrac{4}{11}, \tfrac{6}{11}, \tfrac{8}{11}, \tfrac{10}{11}, \tfrac{12}{11} \},&
         &\{ \tfrac{4}{11}, \tfrac{6}{11}, \tfrac{8}{11}, \tfrac{10}{11}, \tfrac{14}{11} \},&
         &\{ \tfrac{3}{8}, \tfrac{1}{2}, \tfrac{5}{8}, \tfrac{3}{4}, 1 \},\\
         &\{ \tfrac{3}{8}, \tfrac{1}{2}, \tfrac{3}{4}, 1, \tfrac{5}{4} \},&
         &\{ \tfrac{2}{5}, \tfrac{2}{5}, \tfrac{2}{5}, \tfrac{4}{5}, \tfrac{6}{5} \},&
         &\{ \tfrac{2}{5}, \tfrac{2}{5}, -\tfrac{2}{5}, \tfrac{4}{5}, \tfrac{6}{5} \},&
         &\{ \tfrac{2}{5}, \tfrac{2}{5}, \tfrac{3}{5}, \tfrac{4}{5}, 1 \},\\
         &\{ \tfrac{2}{5}, \tfrac{2}{5}, \tfrac{3}{5}, \tfrac{4}{5}, \tfrac{6}{5} \},&
         &\{ \tfrac{2}{5}, \tfrac{2}{5}, \tfrac{4}{5}, \tfrac{4}{5}, \tfrac{6}{5} \},&
         &\{ \tfrac{2}{5}, \tfrac{2}{5}, \tfrac{4}{5}, \tfrac{6}{5}, \tfrac{6}{5} \},&
         &\{ \tfrac{2}{5}, \tfrac{2}{5}, \tfrac{4}{5}, \tfrac{6}{5}, \tfrac{8}{5} \},\\
         &\{ \tfrac{2}{5}, -\tfrac{2}{5}, -\tfrac{2}{5}, \tfrac{4}{5}, \tfrac{6}{5} \},&
         &\{ \tfrac{2}{5}, -\tfrac{2}{5}, \tfrac{3}{5}, \tfrac{4}{5}, \tfrac{6}{5} \},&
         &\{ \tfrac{2}{5}, -\tfrac{2}{5}, \tfrac{4}{5}, \tfrac{4}{5}, \tfrac{6}{5} \},&
         &\{ \tfrac{2}{5}, -\tfrac{2}{5}, \tfrac{4}{5}, \tfrac{6}{5}, \tfrac{6}{5} \},\\
         &\{ \tfrac{2}{5}, -\tfrac{2}{5}, \tfrac{4}{5}, \tfrac{6}{5}, \tfrac{8}{5} \},&
         &\{ \tfrac{2}{5}, -\tfrac{2}{5}, \tfrac{4}{5}, \tfrac{6}{5}, \tfrac{12}{5} \},&
         &\{ \tfrac{2}{5}, -\tfrac{2}{5}, \tfrac{4}{5}, \tfrac{6}{5}, \tfrac{14}{5} \},&
         &\{ \tfrac{2}{5}, \tfrac{1}{2}, \tfrac{3}{5}, \tfrac{4}{5}, 1 \},\\
         &\{ \tfrac{2}{5}, \tfrac{8}{15}, \tfrac{2}{3}, \tfrac{4}{5}, \tfrac{6}{5} \},&
         &\{ \tfrac{2}{5}, \tfrac{8}{15}, \tfrac{2}{3}, \tfrac{14}{15}, \tfrac{16}{15} \},&
         &\{ \tfrac{2}{5}, \tfrac{3}{5}, \tfrac{3}{5}, \tfrac{4}{5}, 1 \},&
         &\{ \tfrac{2}{5}, \tfrac{3}{5}, \tfrac{3}{5}, \tfrac{4}{5}, \tfrac{6}{5} \},\\
         &\{ \tfrac{2}{5}, \tfrac{3}{5}, \tfrac{7}{10}, \tfrac{4}{5}, 1 \},&
         &\{ \tfrac{2}{5}, \tfrac{3}{5}, \tfrac{7}{10}, \tfrac{4}{5}, \tfrac{6}{5} \},&
         &\{ \tfrac{2}{5}, \tfrac{3}{5}, \tfrac{4}{5}, \tfrac{4}{5}, 1 \},&
         &\{ \tfrac{2}{5}, \tfrac{3}{5}, \tfrac{4}{5}, \tfrac{4}{5}, \tfrac{6}{5} \},\\
         &\{ \tfrac{2}{5}, \tfrac{3}{5}, \tfrac{4}{5}, 1, 1 \},&
         &\{ \tfrac{2}{5}, \tfrac{3}{5}, \tfrac{4}{5}, 1, \tfrac{6}{5} \},&
         &\{ \tfrac{2}{5}, \tfrac{3}{5}, \tfrac{4}{5}, 1, \tfrac{7}{5} \},&
         &\{ \tfrac{2}{5}, \tfrac{3}{5}, \tfrac{4}{5}, 1, \tfrac{8}{5} \},\\
         &\{ \tfrac{2}{5}, \tfrac{3}{5}, \tfrac{4}{5}, \tfrac{6}{5}, \tfrac{6}{5} \},&
         &\{ \tfrac{2}{5}, \tfrac{3}{5}, \tfrac{4}{5}, \tfrac{6}{5}, \tfrac{7}{5} \},&
         &\{ \tfrac{2}{5}, \tfrac{3}{5}, \tfrac{4}{5}, \tfrac{6}{5}, \tfrac{8}{5} \},&
         &\{ \tfrac{2}{5}, \tfrac{2}{3}, \tfrac{4}{5}, \tfrac{14}{15}, \tfrac{6}{5} \},\\
         &\{ \tfrac{2}{5}, \tfrac{4}{5}, \tfrac{4}{5}, \tfrac{4}{5}, \tfrac{6}{5} \},&
         &\{ \tfrac{2}{5}, \tfrac{4}{5}, \tfrac{4}{5}, \tfrac{6}{5}, \tfrac{6}{5} \},&
         &\{ \tfrac{2}{5}, \tfrac{4}{5}, \tfrac{4}{5}, \tfrac{6}{5}, \tfrac{8}{5} \},&
         &\{ \tfrac{2}{5}, \tfrac{4}{5}, -\tfrac{4}{5}, \tfrac{6}{5}, \tfrac{8}{5} \},\\
         &\{ \tfrac{2}{5}, \tfrac{4}{5}, \tfrac{6}{5}, \tfrac{6}{5}, \tfrac{6}{5} \},&
         &\{ \tfrac{2}{5}, \tfrac{4}{5}, \tfrac{6}{5}, \tfrac{6}{5}, \tfrac{8}{5} \},&
         &\{ \tfrac{2}{5}, \tfrac{4}{5}, \tfrac{6}{5}, -\tfrac{6}{5}, \tfrac{8}{5} \},&
         &\{ \tfrac{2}{5}, \tfrac{4}{5}, \tfrac{6}{5}, \tfrac{8}{5}, \tfrac{8}{5} \},\\
         &\{ \tfrac{3}{7}, \tfrac{4}{7}, \tfrac{5}{7}, \tfrac{6}{7}, 1 \},&
         &\{ \tfrac{3}{7}, \tfrac{4}{7}, \tfrac{5}{7}, \tfrac{6}{7}, \tfrac{8}{7} \},&
         &\{ \tfrac{4}{9}, \tfrac{4}{9}, \tfrac{2}{3}, \tfrac{8}{9}, \tfrac{10}{9} \},&
         &\{ \tfrac{4}{9}, \tfrac{5}{9}, \tfrac{2}{3}, \tfrac{7}{9}, 1 \},\\
         &\{ \tfrac{4}{9}, \tfrac{5}{9}, \tfrac{2}{3}, \tfrac{8}{9}, 1 \},&
         &\{ \tfrac{4}{9}, \tfrac{5}{9}, \tfrac{2}{3}, \tfrac{8}{9}, \tfrac{10}{9} \},&
         &\{ \tfrac{4}{9}, \tfrac{2}{3}, \tfrac{2}{3}, \tfrac{8}{9}, \tfrac{10}{9} \},&
         &\{ \tfrac{4}{9}, \tfrac{2}{3}, \tfrac{7}{9}, \tfrac{8}{9}, \tfrac{10}{9} \},\\
         &\{ \tfrac{4}{9}, \tfrac{2}{3}, \tfrac{8}{9}, \tfrac{8}{9}, \tfrac{10}{9} \},&
         &\{ \tfrac{4}{9}, \tfrac{2}{3}, \tfrac{8}{9}, \tfrac{10}{9}, \tfrac{10}{9} \},&
         &\{ \tfrac{4}{9}, \tfrac{2}{3}, \tfrac{8}{9}, \tfrac{10}{9}, \tfrac{4}{3} \},&
         &\{ \tfrac{4}{9}, \tfrac{2}{3}, \tfrac{8}{9}, \tfrac{10}{9}, \tfrac{14}{9} \},\\
         &\{ \tfrac{6}{13}, \tfrac{8}{13}, \tfrac{10}{13}, \tfrac{12}{13}, \tfrac{14}{13} \},&
         &\{ \tfrac{1}{2}, \tfrac{1}{2}, \tfrac{1}{2}, \tfrac{1}{2}, 1 \},&
         &\{ \tfrac{1}{2}, \tfrac{1}{2}, \tfrac{1}{2}, \tfrac{3}{4}, 1 \},&
         &\{ \tfrac{1}{2}, \tfrac{1}{2}, \tfrac{1}{2}, 1, 1 \},\\
         &\{ \tfrac{1}{2}, \tfrac{1}{2}, \tfrac{1}{2}, 1, \tfrac{3}{2} \},&
         &\{ \tfrac{1}{2}, \tfrac{1}{2}, -\tfrac{1}{2}, 1, \tfrac{3}{2} \},&
         &\{ \tfrac{1}{2}, \tfrac{1}{2}, \tfrac{5}{8}, \tfrac{3}{4}, 1 \},&
         &\{ \tfrac{1}{2}, \tfrac{1}{2}, \tfrac{2}{3}, \tfrac{5}{6}, 1 \},\\
         &\{ \tfrac{1}{2}, \tfrac{1}{2}, \tfrac{3}{4}, \tfrac{3}{4}, 1 \},&
         &\{ \tfrac{1}{2}, \tfrac{1}{2}, \tfrac{3}{4}, 1, 1 \},&
         &\{ \tfrac{1}{2}, \tfrac{1}{2}, \tfrac{3}{4}, 1, \tfrac{5}{4} \},&
         &\{ \tfrac{1}{2}, \tfrac{1}{2}, \tfrac{3}{4}, 1, \tfrac{3}{2} \},\\
         &\{ \tfrac{1}{2}, \tfrac{1}{2}, 1, 1, 1 \},&
         &\{ \tfrac{1}{2}, \tfrac{1}{2}, 1, 1, \tfrac{3}{2} \},&
         &\{ \tfrac{1}{2}, \tfrac{1}{2}, 1, -1, \tfrac{3}{2} \},&
         &\{ \tfrac{1}{2}, \tfrac{1}{2}, 1, \tfrac{3}{2}, \tfrac{3}{2} \},\\
         &\{ \tfrac{1}{2}, -\tfrac{1}{2}, -\tfrac{1}{2}, 1, \tfrac{3}{2} \},&
         &\{ \tfrac{1}{2}, -\tfrac{1}{2}, \tfrac{3}{4}, 1, \tfrac{5}{4} \},&
         &\{ \tfrac{1}{2}, -\tfrac{1}{2}, \tfrac{3}{4}, 1, \tfrac{3}{2} \},&
         &\{ \tfrac{1}{2}, -\tfrac{1}{2}, 1, 1, \tfrac{3}{2} \},\\
         &\{ \tfrac{1}{2}, -\tfrac{1}{2}, 1, -1, \tfrac{3}{2} \},&
         &\{ \tfrac{1}{2}, -\tfrac{1}{2}, 1, \tfrac{5}{4}, \tfrac{3}{2} \},&
         &\{ \tfrac{1}{2}, -\tfrac{1}{2}, 1, \tfrac{3}{2}, \tfrac{3}{2} \},&
         &\{ \tfrac{1}{2}, -\tfrac{1}{2}, 1, \tfrac{3}{2}, \tfrac{5}{2} \},\\
         &\{ \tfrac{1}{2}, -\tfrac{1}{2}, 1, \tfrac{3}{2}, 3 \},&
         &\{ \tfrac{1}{2}, \tfrac{7}{12}, \tfrac{2}{3}, \tfrac{3}{4}, 1 \},&
         &\{ \tfrac{1}{2}, \tfrac{7}{12}, \tfrac{2}{3}, \tfrac{5}{6}, 1 \},&
         &\{ \tfrac{1}{2}, \tfrac{3}{5}, \tfrac{7}{10}, \tfrac{4}{5}, 1 \},\\
         &\{ \tfrac{1}{2}, \tfrac{5}{8}, \tfrac{5}{8}, \tfrac{3}{4}, 1 \},&
         &\{ \tfrac{1}{2}, \tfrac{5}{8}, \tfrac{11}{16}, \tfrac{3}{4}, 1 \},&
         &\{ \tfrac{1}{2}, \tfrac{5}{8}, \tfrac{3}{4}, \tfrac{3}{4}, 1 \},&
         &\{ \tfrac{1}{2}, \tfrac{5}{8}, \tfrac{3}{4}, \tfrac{7}{8}, 1 \},\\
         &\{ \tfrac{1}{2}, \tfrac{5}{8}, \tfrac{3}{4}, 1, 1 \},&
         &\{ \tfrac{1}{2}, \tfrac{5}{8}, \tfrac{3}{4}, 1, \tfrac{5}{4} \},&
         &\{ \tfrac{1}{2}, \tfrac{5}{8}, \tfrac{3}{4}, 1, \tfrac{11}{8} \},&
         &\{ \tfrac{1}{2}, \tfrac{5}{8}, \tfrac{3}{4}, 1, \tfrac{3}{2} \},\\
         &\{ \tfrac{1}{2}, \tfrac{2}{3}, \tfrac{2}{3}, \tfrac{5}{6}, 1 \},&
         &\{ \tfrac{1}{2}, \tfrac{2}{3}, \tfrac{3}{4}, \tfrac{5}{6}, 1 \},&
         &\{ \tfrac{1}{2}, \tfrac{2}{3}, \tfrac{5}{6}, \tfrac{5}{6}, 1 \},&
         &\{ \tfrac{1}{2}, \tfrac{2}{3}, \tfrac{5}{6}, 1, 1 \},\\
         &\{ \tfrac{1}{2}, \tfrac{2}{3}, \tfrac{5}{6}, 1, \tfrac{7}{6} \},&
         &\{ \tfrac{1}{2}, \tfrac{2}{3}, \tfrac{5}{6}, 1, \tfrac{4}{3} \},&
         &\{ \tfrac{1}{2}, \tfrac{2}{3}, \tfrac{5}{6}, 1, \tfrac{3}{2} \},&
         &\{ \tfrac{1}{2}, \tfrac{3}{4}, \tfrac{3}{4}, \tfrac{3}{4}, 1 \},\\
         &\{ \tfrac{1}{2}, \tfrac{3}{4}, \tfrac{3}{4}, 1, 1 \},&
         &\{ \tfrac{1}{2}, \tfrac{3}{4}, \tfrac{3}{4}, 1, \tfrac{5}{4} \},&
         &\{ \tfrac{1}{2}, \tfrac{3}{4}, \tfrac{3}{4}, 1, \tfrac{3}{2} \},&
         &\{ \tfrac{1}{2}, \tfrac{3}{4}, 1, 1, 1 \},\\
         &\{ \tfrac{1}{2}, \tfrac{3}{4}, 1, 1, \tfrac{5}{4} \},&
         &\{ \tfrac{1}{2}, \tfrac{3}{4}, 1, 1, \tfrac{3}{2} \},&
         &\{ \tfrac{1}{2}, \tfrac{3}{4}, 1, -1, \tfrac{3}{2} \},&
         &\{ \tfrac{1}{2}, \tfrac{3}{4}, 1, \tfrac{5}{4}, \tfrac{5}{4} \},\\
         &\{ \tfrac{1}{2}, \tfrac{3}{4}, 1, \tfrac{5}{4}, \tfrac{3}{2} \},&
         &\{ \tfrac{1}{2}, \tfrac{3}{4}, 1, \tfrac{3}{2}, \tfrac{3}{2} \},&
         &\{ \tfrac{1}{2}, 1, 1, 1, 1 \},&
         &\{ \tfrac{1}{2}, 1, 1, 1, \tfrac{3}{2} \},\\
         &\{ \tfrac{1}{2}, 1, 1, -1, \tfrac{3}{2} \},&
         &\{ \tfrac{1}{2}, 1, 1, \tfrac{3}{2}, \tfrac{3}{2} \},&
         &\{ \tfrac{1}{2}, 1, -1, -1, \tfrac{3}{2} \},&
         &\{ \tfrac{1}{2}, 1, -1, \tfrac{3}{2}, \tfrac{3}{2} \},\\
         &\{ \tfrac{1}{2}, 1, -1, \tfrac{3}{2}, \tfrac{5}{2} \},&
         &\{ \tfrac{1}{2}, 1, -1, \tfrac{3}{2}, 3 \},&
         &\{ \tfrac{1}{2}, 1, -1, \tfrac{3}{2}, 4 \},&
         &\{ \tfrac{1}{2}, 1, \tfrac{3}{2}, \tfrac{3}{2}, \tfrac{3}{2} \},\\
         &\{ \tfrac{2}{3}, \tfrac{2}{3}, \tfrac{2}{3}, \tfrac{2}{3}, \tfrac{2}{3} \},&
         &\{ \tfrac{2}{3}, \tfrac{2}{3}, \tfrac{2}{3}, \tfrac{2}{3}, \tfrac{4}{3} \},&
         &\{ \tfrac{2}{3}, \tfrac{2}{3}, \tfrac{2}{3}, -\tfrac{2}{3}, \tfrac{4}{3} \},&
         &\{ \tfrac{2}{3}, \tfrac{2}{3}, \tfrac{2}{3}, \tfrac{4}{3}, \tfrac{4}{3} \},\\
         &\{ \tfrac{2}{3}, \tfrac{2}{3}, -\tfrac{2}{3}, -\tfrac{2}{3}, \tfrac{4}{3} \},&
         &\{ \tfrac{2}{3}, \tfrac{2}{3}, -\tfrac{2}{3}, \tfrac{4}{3}, \tfrac{4}{3} \},&
         &\{ \tfrac{2}{3}, \tfrac{2}{3}, -\tfrac{2}{3}, \tfrac{4}{3}, \tfrac{8}{3} \},&
         &\{ \tfrac{2}{3}, \tfrac{2}{3}, -\tfrac{2}{3}, \tfrac{4}{3}, \tfrac{10}{3} \},\\
         &\{ \tfrac{2}{3}, \tfrac{2}{3}, \tfrac{4}{3}, \tfrac{4}{3}, \tfrac{4}{3} \},&
         &\{ \tfrac{2}{3}, -\tfrac{2}{3}, -\tfrac{2}{3}, -\tfrac{2}{3}, \tfrac{4}{3} \},&
         &\{ \tfrac{2}{3}, -\tfrac{2}{3}, -\tfrac{2}{3}, \tfrac{4}{3}, \tfrac{4}{3} \},&
         &\{ \tfrac{2}{3}, -\tfrac{2}{3}, -\tfrac{2}{3}, \tfrac{4}{3}, \tfrac{8}{3} \},\\
         &\{ \tfrac{2}{3}, -\tfrac{2}{3}, -\tfrac{2}{3}, \tfrac{4}{3}, \tfrac{10}{3} \},&
         &\{ \tfrac{2}{3}, -\tfrac{2}{3}, 1, \tfrac{4}{3}, \tfrac{5}{3} \},&
         &\{ \tfrac{2}{3}, -\tfrac{2}{3}, \tfrac{4}{3}, \tfrac{4}{3}, \tfrac{4}{3} \},&
         &\{ \tfrac{2}{3}, -\tfrac{2}{3}, \tfrac{4}{3}, \tfrac{4}{3}, \tfrac{8}{3} \},\\
         &\{ \tfrac{2}{3}, -\tfrac{2}{3}, \tfrac{4}{3}, \tfrac{4}{3}, \tfrac{10}{3} \},&
         &\{ \tfrac{2}{3}, -\tfrac{2}{3}, \tfrac{4}{3}, -\tfrac{4}{3}, \tfrac{8}{3} \},&
         &\{ \tfrac{2}{3}, -\tfrac{2}{3}, \tfrac{4}{3}, -\tfrac{4}{3}, \tfrac{10}{3} \},&
         &\{ \tfrac{2}{3}, -\tfrac{2}{3}, \tfrac{4}{3}, \tfrac{8}{3}, \tfrac{8}{3} \},\\
         &\{ \tfrac{2}{3}, -\tfrac{2}{3}, \tfrac{4}{3}, \tfrac{8}{3}, \tfrac{10}{3} \},&
         &\{ \tfrac{2}{3}, -\tfrac{2}{3}, \tfrac{4}{3}, \tfrac{8}{3}, -\tfrac{10}{3} \},&
         &\{ \tfrac{2}{3}, -\tfrac{2}{3}, \tfrac{4}{3}, -\tfrac{8}{3}, \tfrac{10}{3} \},&
         &\{ \tfrac{2}{3}, -\tfrac{2}{3}, \tfrac{4}{3}, \tfrac{10}{3}, \tfrac{10}{3} \},\\
         &\{ \tfrac{2}{3}, -\tfrac{2}{3}, \tfrac{4}{3}, \tfrac{10}{3}, -\tfrac{14}{3} \},&
         &\{ \tfrac{2}{3}, -\tfrac{2}{3}, -\tfrac{4}{3}, \tfrac{8}{3}, \tfrac{10}{3} \},&
         &\{ \tfrac{2}{3}, \tfrac{4}{3}, \tfrac{4}{3}, \tfrac{4}{3}, \tfrac{4}{3} \},&
         &\{ 1, 1, 1, 1, 1 \}; \refstepcounter{equation} \tag{\theequation}
\end{align*}
$237$ of them have $N_2 > N_0$ and no SUSY solution:
\begin{align*}
\hat r = \mbox{}
         &\{ 2, 2, 2, 2, 0 \},&
         &\{ 2, 2, 2, 2, -2 \},&
         &\{ 2, 2, 2, 0, 0 \},&
         &\{ 2, 2, 2, 0, -2 \},\\
         &\{ 2, 2, 2, 0, 1 \},&
         &\{ 2, 2, 2, -2, -2 \},&
         &\{ 2, 2, 2, -2, 4 \},&
         &\{ 2, 2, 2, -2, 6 \},\\
         &\{ 2, 2, 2, \tfrac{2}{3}, -\tfrac{2}{3} \},&
         &\{ 2, 2, 2, 1, -1 \},&
         &\{ 2, 2, 0, -2, -2 \},&
         &\{ 2, 2, 0, -2, 1 \},\\
         &\{ 2, 2, 0, -2, 4 \},&
         &\{ 2, 2, 0, -2, 6 \},&
         &\{ 2, 2, 0, \tfrac{1}{2}, 1 \},&
         &\{ 2, 2, 0, \tfrac{2}{3}, -\tfrac{2}{3} \},\\
         &\{ 2, 2, 0, \tfrac{2}{3}, \tfrac{4}{3} \},&
         &\{ 2, 2, 0, 1, 1 \},&
         &\{ 2, 2, 0, 1, -1 \},&
         &\{ 2, 2, -2, -2, -2 \},\\
         &\{ 2, 2, -2, -2, 4 \},&
         &\{ 2, 2, -2, -2, 6 \},&
         &\{ 2, 2, -2, \tfrac{2}{3}, -\tfrac{2}{3} \},&
         &\{ 2, 2, -2, \tfrac{2}{3}, \tfrac{10}{3} \},\\
         &\{ 2, 2, -2, 1, -1 \},&
         &\{ 2, 2, -2, 1, 3 \},&
         &\{ 2, 2, -2, -1, 4 \},&
         &\{ 2, 2, -2, 4, 4 \},\\
         &\{ 2, 2, -2, 4, -4 \},&
         &\{ 2, 2, -2, 4, 6 \},&
         &\{ 2, 2, -2, 4, -6 \},&
         &\{ 2, 2, -2, -4, 6 \},\\
         &\{ 2, 2, -2, 6, 6 \},&
         &\{ 2, 2, -2, 6, -6 \},&
         &\{ 2, 2, -2, 6, -10 \},&
         &\{ 2, 2, \tfrac{2}{5}, -\tfrac{2}{5}, \tfrac{6}{5} \},\\
         &\{ 2, 2, \tfrac{1}{2}, -\tfrac{1}{2}, 1 \},&
         &\{ 2, 2, \tfrac{1}{2}, 1, -1 \},&
         &\{ 2, 2, \tfrac{2}{3}, \tfrac{2}{3}, -\tfrac{2}{3} \},&
         &\{ 2, 2, \tfrac{2}{3}, -\tfrac{2}{3}, -\tfrac{2}{3} \},\\
         &\{ 2, 2, \tfrac{2}{3}, -\tfrac{2}{3}, \tfrac{4}{3} \},&
         &\{ 2, 2, \tfrac{2}{3}, -\tfrac{2}{3}, \tfrac{8}{3} \},&
         &\{ 2, 2, \tfrac{2}{3}, -\tfrac{2}{3}, \tfrac{10}{3} \},&
         &\{ 2, 2, \tfrac{2}{3}, \tfrac{4}{3}, -\tfrac{4}{3} \},\\
         &\{ 2, 2, 1, 1, -1 \},&
         &\{ 2, 2, 1, -1, -1 \},&
         &\{ 2, 2, 1, -1, \tfrac{3}{2} \},&
         &\{ 2, 2, 1, -1, 3 \},\\
         &\{ 2, 2, 1, -1, 4 \},&
         &\{ 2, -2, -2, -2, -2 \},&
         &\{ 2, -2, -2, -2, 4 \},&
         &\{ 2, -2, -2, -2, 6 \},\\
         &\{ 2, -2, -2, \tfrac{2}{3}, -\tfrac{2}{3} \},&
         &\{ 2, -2, -2, \tfrac{2}{3}, \tfrac{10}{3} \},&
         &\{ 2, -2, -2, 1, -1 \},&
         &\{ 2, -2, -2, 1, 3 \},\\
         &\{ 2, -2, -2, -1, 4 \},&
         &\{ 2, -2, -2, 4, 4 \},&
         &\{ 2, -2, -2, 4, -4 \},&
         &\{ 2, -2, -2, 4, 6 \},\\
         &\{ 2, -2, -2, 4, -6 \},&
         &\{ 2, -2, -2, -4, 6 \},&
         &\{ 2, -2, -2, 6, 6 \},&
         &\{ 2, -2, -2, 6, -10 \},\\
         &\{ 2, -2, \tfrac{2}{5}, -\tfrac{2}{5}, \tfrac{6}{5} \},&
         &\{ 2, -2, -\tfrac{2}{5}, \tfrac{6}{5}, \tfrac{14}{5} \},&
         &\{ 2, -2, \tfrac{1}{2}, -\tfrac{1}{2}, 1 \},&
         &\{ 2, -2, \tfrac{1}{2}, 1, -1 \},\\
         &\{ 2, -2, \tfrac{1}{2}, 1, 3 \},&
         &\{ 2, -2, \tfrac{1}{2}, 1, \tfrac{7}{2} \},&
         &\{ 2, -2, -\tfrac{1}{2}, 1, 3 \},&
         &\{ 2, -2, \tfrac{2}{3}, \tfrac{2}{3}, -\tfrac{2}{3} \},\\
         &\{ 2, -2, \tfrac{2}{3}, \tfrac{2}{3}, \tfrac{10}{3} \},&
         &\{ 2, -2, \tfrac{2}{3}, -\tfrac{2}{3}, -\tfrac{2}{3} \},&
         &\{ 2, -2, \tfrac{2}{3}, -\tfrac{2}{3}, \tfrac{4}{3} \},&
         &\{ 2, -2, \tfrac{2}{3}, -\tfrac{2}{3}, \tfrac{8}{3} \},\\
         &\{ 2, -2, \tfrac{2}{3}, -\tfrac{2}{3}, \tfrac{10}{3} \},&
         &\{ 2, -2, \tfrac{2}{3}, -\tfrac{2}{3}, 4 \},&
         &\{ 2, -2, \tfrac{2}{3}, -\tfrac{2}{3}, \tfrac{14}{3} \},&
         &\{ 2, -2, \tfrac{2}{3}, -\tfrac{2}{3}, 6 \},\\
         &\{ 2, -2, \tfrac{2}{3}, \tfrac{4}{3}, -\tfrac{4}{3} \},&
         &\{ 2, -2, \tfrac{2}{3}, \tfrac{4}{3}, \tfrac{8}{3} \},&
         &\{ 2, -2, \tfrac{2}{3}, \tfrac{4}{3}, \tfrac{10}{3} \},&
         &\{ 2, -2, \tfrac{2}{3}, -\tfrac{4}{3}, \tfrac{10}{3} \},\\
         &\{ 2, -2, \tfrac{2}{3}, -\tfrac{8}{3}, 4 \},&
         &\{ 2, -2, \tfrac{2}{3}, \tfrac{10}{3}, \tfrac{10}{3} \},&
         &\{ 2, -2, \tfrac{2}{3}, \tfrac{10}{3}, 4 \},&
         &\{ 2, -2, \tfrac{2}{3}, \tfrac{10}{3}, -\tfrac{14}{3} \},\\
         &\{ 2, -2, \tfrac{2}{3}, \tfrac{10}{3}, 6 \},&
         &\{ 2, -2, \tfrac{2}{3}, -\tfrac{14}{3}, 6 \},&
         &\{ 2, -2, -\tfrac{2}{3}, \tfrac{4}{3}, \tfrac{8}{3} \},&
         &\{ 2, -2, 1, 1, -1 \},\\
         &\{ 2, -2, 1, 1, 3 \},&
         &\{ 2, -2, 1, -1, -1 \},&
         &\{ 2, -2, 1, -1, \tfrac{3}{2} \},&
         &\{ 2, -2, 1, -1, 3 \},\\
         &\{ 2, -2, 1, -1, 4 \},&
         &\{ 2, -2, 1, -1, 5 \},&
         &\{ 2, -2, 1, -1, 6 \},&
         &\{ 2, -2, 1, 3, 3 \},\\
         &\{ 2, -2, 1, 3, 4 \},&
         &\{ 2, -2, 1, 3, -4 \},&
         &\{ 2, -2, 1, 3, 6 \},&
         &\{ 2, -2, 1, -3, 4 \},\\
         &\{ 2, -2, 1, -5, 6 \},&
         &\{ 2, -2, -1, -1, 4 \},&
         &\{ 2, -2, -1, \tfrac{3}{2}, 4 \},&
         &\{ 2, -2, -1, 3, 4 \},\\
         &\{ 2, -2, -1, 4, 4 \},&
         &\{ 2, -2, -1, 4, -4 \},&
         &\{ 2, -2, -1, 4, 5 \},&
         &\{ 2, -2, -1, 4, 6 \},\\
         &\{ 2, -2, -1, 4, -6 \},&
         &\{ 2, -2, 3, 4, -4 \},&
         &\{ 2, -2, 3, -4, 6 \},&
         &\{ 2, -2, 4, 4, 4 \},\\
         &\{ 2, -2, 4, 4, 6 \},&
         &\{ 2, -2, 4, 4, -6 \},&
         &\{ 2, -2, 4, -4, -4 \},&
         &\{ 2, -2, 4, -4, 6 \},\\
         &\{ 2, -2, 4, -4, -6 \},&
         &\{ 2, -2, 4, -4, 10 \},&
         &\{ 2, -2, 4, 6, 6 \},&
         &\{ 2, -2, 4, 6, -8 \},\\
         &\{ 2, -2, 4, 6, -10 \},&
         &\{ 2, -2, 4, -6, -6 \},&
         &\{ 2, -2, 4, -6, 8 \},&
         &\{ 2, -2, 4, -6, 10 \},\\
         &\{ 2, -2, 4, -6, 14 \},&
         &\{ 2, -2, -4, -4, 6 \},&
         &\{ 2, -2, -4, 6, 6 \},&
         &\{ 2, -2, -4, 6, -6 \},\\
         &\{ 2, -2, -4, 6, 8 \},&
         &\{ 2, -2, -4, 6, 10 \},&
         &\{ 2, -2, -4, 6, -10 \},&
         &\{ 2, -2, 6, 6, 6 \},\\
         &\{ 2, -2, 6, 6, -10 \},&
         &\{ 2, -2, 6, -6, 8 \},&
         &\{ 2, -2, 6, -6, -10 \},&
         &\{ 2, -2, 6, -6, 14 \},\\
         &\{ 2, -2, 6, -10, -10 \},&
         &\{ 2, -2, 6, -10, 12 \},&
         &\{ 2, -2, 6, -10, 14 \},&
         &\{ 2, -2, 6, -10, 22 \},\\
         &\{ 2, \tfrac{2}{9}, -\tfrac{2}{9}, \tfrac{2}{3}, \tfrac{14}{9} \},&
         &\{ 2, \tfrac{1}{4}, 1, -1, \tfrac{3}{2} \},&
         &\{ 2, \tfrac{2}{7}, -\tfrac{2}{7}, \tfrac{6}{7}, \tfrac{10}{7} \},&
         &\{ 2, \tfrac{2}{7}, \tfrac{6}{7}, -\tfrac{6}{7}, \tfrac{10}{7} \},\\
         &\{ 2, \tfrac{1}{3}, -\tfrac{1}{3}, \tfrac{2}{3}, \tfrac{4}{3} \},&
         &\{ 2, \tfrac{1}{3}, -\tfrac{1}{3}, \tfrac{2}{3}, \tfrac{5}{3} \},&
         &\{ 2, \tfrac{1}{3}, -\tfrac{1}{3}, 1, \tfrac{4}{3} \},&
         &\{ 2, \tfrac{1}{3}, \tfrac{2}{3}, -\tfrac{2}{3}, 1 \},\\
         &\{ 2, \tfrac{1}{3}, \tfrac{2}{3}, -\tfrac{2}{3}, \tfrac{4}{3} \},&
         &\{ 2, \tfrac{1}{3}, \tfrac{2}{3}, 1, -1 \},&
         &\{ 2, \tfrac{1}{3}, \tfrac{2}{3}, \tfrac{4}{3}, -\tfrac{4}{3} \},&
         &\{ 2, -\tfrac{1}{3}, \tfrac{2}{3}, -\tfrac{2}{3}, \tfrac{8}{3} \},\\
         &\{ 2, \tfrac{2}{5}, -\tfrac{2}{5}, -\tfrac{2}{5}, \tfrac{6}{5} \},&
         &\{ 2, \tfrac{2}{5}, -\tfrac{2}{5}, \tfrac{4}{5}, \tfrac{6}{5} \},&
         &\{ 2, \tfrac{2}{5}, -\tfrac{2}{5}, \tfrac{4}{5}, \tfrac{8}{5} \},&
         &\{ 2, \tfrac{2}{5}, -\tfrac{2}{5}, \tfrac{6}{5}, \tfrac{6}{5} \},\\
         &\{ 2, \tfrac{2}{5}, -\tfrac{2}{5}, \tfrac{6}{5}, -\tfrac{6}{5} \},&
         &\{ 2, \tfrac{2}{5}, -\tfrac{2}{5}, \tfrac{6}{5}, \tfrac{8}{5} \},&
         &\{ 2, \tfrac{2}{5}, -\tfrac{2}{5}, \tfrac{6}{5}, \tfrac{12}{5} \},&
         &\{ 2, \tfrac{2}{5}, -\tfrac{2}{5}, \tfrac{6}{5}, \tfrac{14}{5} \},\\
         &\{ 2, \tfrac{2}{5}, \tfrac{4}{5}, -\tfrac{4}{5}, \tfrac{6}{5} \},&
         &\{ 2, \tfrac{2}{5}, \tfrac{4}{5}, \tfrac{6}{5}, -\tfrac{6}{5} \},&
         &\{ 2, \tfrac{1}{2}, \tfrac{1}{2}, 1, -1 \},&
         &\{ 2, \tfrac{1}{2}, -\tfrac{1}{2}, -\tfrac{1}{2}, 1 \},\\
         &\{ 2, \tfrac{1}{2}, -\tfrac{1}{2}, \tfrac{3}{4}, 1 \},&
         &\{ 2, \tfrac{1}{2}, -\tfrac{1}{2}, 1, 1 \},&
         &\{ 2, \tfrac{1}{2}, -\tfrac{1}{2}, 1, -1 \},&
         &\{ 2, \tfrac{1}{2}, -\tfrac{1}{2}, 1, \tfrac{5}{4} \},\\
         &\{ 2, \tfrac{1}{2}, -\tfrac{1}{2}, 1, \tfrac{3}{2} \},&
         &\{ 2, \tfrac{1}{2}, -\tfrac{1}{2}, 1, \tfrac{5}{2} \},&
         &\{ 2, \tfrac{1}{2}, -\tfrac{1}{2}, 1, 3 \},&
         &\{ 2, \tfrac{1}{2}, \tfrac{3}{4}, -\tfrac{3}{4}, 1 \},\\
         &\{ 2, \tfrac{1}{2}, \tfrac{3}{4}, 1, -1 \},&
         &\{ 2, \tfrac{1}{2}, 1, 1, -1 \},&
         &\{ 2, \tfrac{1}{2}, 1, -1, -1 \},&
         &\{ 2, \tfrac{1}{2}, 1, -1, \tfrac{3}{2} \},\\
         &\{ 2, \tfrac{1}{2}, 1, -1, \tfrac{5}{2} \},&
         &\{ 2, \tfrac{1}{2}, 1, -1, 3 \},&
         &\{ 2, \tfrac{1}{2}, 1, -1, 4 \},&
         &\{ 2, \tfrac{1}{2}, 1, \tfrac{3}{2}, -\tfrac{3}{2} \},\\
         &\{ 2, -\tfrac{1}{2}, 1, -1, \tfrac{3}{2} \},&
         &\{ 2, -\tfrac{1}{2}, 1, -1, 3 \},&
         &\{ 2, \tfrac{2}{3}, \tfrac{2}{3}, \tfrac{2}{3}, -\tfrac{2}{3} \},&
         &\{ 2, \tfrac{2}{3}, \tfrac{2}{3}, -\tfrac{2}{3}, -\tfrac{2}{3} \},\\
         &\{ 2, \tfrac{2}{3}, \tfrac{2}{3}, -\tfrac{2}{3}, \tfrac{4}{3} \},&
         &\{ 2, \tfrac{2}{3}, \tfrac{2}{3}, -\tfrac{2}{3}, \tfrac{8}{3} \},&
         &\{ 2, \tfrac{2}{3}, \tfrac{2}{3}, -\tfrac{2}{3}, \tfrac{10}{3} \},&
         &\{ 2, \tfrac{2}{3}, \tfrac{2}{3}, \tfrac{4}{3}, -\tfrac{4}{3} \},\\
         &\{ 2, \tfrac{2}{3}, -\tfrac{2}{3}, -\tfrac{2}{3}, -\tfrac{2}{3} \},&
         &\{ 2, \tfrac{2}{3}, -\tfrac{2}{3}, -\tfrac{2}{3}, \tfrac{4}{3} \},&
         &\{ 2, \tfrac{2}{3}, -\tfrac{2}{3}, -\tfrac{2}{3}, \tfrac{8}{3} \},&
         &\{ 2, \tfrac{2}{3}, -\tfrac{2}{3}, -\tfrac{2}{3}, \tfrac{10}{3} \},\\
         &\{ 2, \tfrac{2}{3}, -\tfrac{2}{3}, 1, -1 \},&
         &\{ 2, \tfrac{2}{3}, -\tfrac{2}{3}, 1, \tfrac{5}{3} \},&
         &\{ 2, \tfrac{2}{3}, -\tfrac{2}{3}, \tfrac{4}{3}, \tfrac{4}{3} \},&
         &\{ 2, \tfrac{2}{3}, -\tfrac{2}{3}, \tfrac{4}{3}, -\tfrac{4}{3} \},\\
         &\{ 2, \tfrac{2}{3}, -\tfrac{2}{3}, \tfrac{4}{3}, \tfrac{8}{3} \},&
         &\{ 2, \tfrac{2}{3}, -\tfrac{2}{3}, \tfrac{4}{3}, \tfrac{10}{3} \},&
         &\{ 2, \tfrac{2}{3}, -\tfrac{2}{3}, -\tfrac{4}{3}, \tfrac{8}{3} \},&
         &\{ 2, \tfrac{2}{3}, -\tfrac{2}{3}, -\tfrac{4}{3}, \tfrac{10}{3} \},\\
         &\{ 2, \tfrac{2}{3}, -\tfrac{2}{3}, \tfrac{8}{3}, \tfrac{8}{3} \},&
         &\{ 2, \tfrac{2}{3}, -\tfrac{2}{3}, \tfrac{8}{3}, -\tfrac{8}{3} \},&
         &\{ 2, \tfrac{2}{3}, -\tfrac{2}{3}, \tfrac{8}{3}, \tfrac{10}{3} \},&
         &\{ 2, \tfrac{2}{3}, -\tfrac{2}{3}, \tfrac{8}{3}, -\tfrac{10}{3} \},\\
         &\{ 2, \tfrac{2}{3}, -\tfrac{2}{3}, \tfrac{10}{3}, \tfrac{10}{3} \},&
         &\{ 2, \tfrac{2}{3}, -\tfrac{2}{3}, \tfrac{10}{3}, -\tfrac{14}{3} \},&
         &\{ 2, \tfrac{2}{3}, 1, -1, \tfrac{7}{3} \},&
         &\{ 2, \tfrac{2}{3}, \tfrac{4}{3}, -\tfrac{4}{3}, -\tfrac{4}{3} \},\\
         &\{ 2, \tfrac{2}{3}, \tfrac{4}{3}, -\tfrac{4}{3}, \tfrac{5}{3} \},&
         &\{ 2, \tfrac{2}{3}, \tfrac{4}{3}, -\tfrac{4}{3}, \tfrac{10}{3} \},&
         &\{ 2, \tfrac{2}{3}, \tfrac{4}{3}, -\tfrac{4}{3}, \tfrac{14}{3} \},&
         &\{ 2, -\tfrac{2}{3}, \tfrac{4}{3}, -\tfrac{4}{3}, \tfrac{10}{3} \},\\
         &\{ 2, 1, 1, 1, -1 \},&
         &\{ 2, 1, 1, -1, -1 \},&
         &\{ 2, 1, 1, -1, \tfrac{3}{2} \},&
         &\{ 2, 1, 1, -1, 3 \},\\
         &\{ 2, 1, 1, -1, 4 \},&
         &\{ 2, 1, -1, -1, -1 \},&
         &\{ 2, 1, -1, -1, \tfrac{3}{2} \},&
         &\{ 2, 1, -1, -1, 3 \},\\
         &\{ 2, 1, -1, -1, 4 \},&
         &\{ 2, 1, -1, \tfrac{3}{2}, \tfrac{3}{2} \},&
         &\{ 2, 1, -1, \tfrac{3}{2}, -\tfrac{3}{2} \},&
         &\{ 2, 1, -1, \tfrac{3}{2}, 3 \},\\
         &\{ 2, 1, -1, \tfrac{3}{2}, 4 \},&
         &\{ 2, 1, -1, 3, 3 \},&
         &\{ 2, 1, -1, 3, -3 \},&
         &\{ 2, 1, -1, 3, 4 \},\\
         &\{ 2, 1, -1, 3, -4 \},&
         &\{ 2, 1, -1, -3, 4 \},&
         &\{ 2, 1, -1, 4, 4 \},&
         &\{ 2, 1, -1, 4, -6 \},\\
         &\{ 2, 4, -4, 6, -6 \}; \refstepcounter{equation} \tag{\theequation}
\end{align*}
and $18$ of them have $N_2 > N_0$ and SUSY solutions:
\begin{align*}
\hat r = \mbox{}
         &\{ 2, -2, -2, 6, -6 \},&
         &\{ 2, -2, \tfrac{2}{3}, \tfrac{10}{3}, -\tfrac{10}{3} \},&
         &\{ 2, -2, 1, 3, -3 \},&
         &\{ 2, -2, 4, 4, -4 \},\\
         &\{ 2, -2, 4, -4, 8 \},&
         &\{ 2, -2, 4, 6, -6 \},&
         &\{ 2, -2, 6, 6, -6 \},&
         &\{ 2, -2, 6, -6, -6 \},\\
         &\{ 2, -2, 6, -6, 10 \},&
         &\{ 2, -2, 6, 10, -10 \},&
         &\{ 2, \tfrac{2}{9}, -\tfrac{2}{9}, \tfrac{2}{3}, \tfrac{10}{9} \},&
         &\{ 2, \tfrac{1}{3}, -\tfrac{1}{3}, \tfrac{2}{3}, 1 \},\\
         &\{ 2, \tfrac{2}{5}, \tfrac{2}{5}, -\tfrac{2}{5}, \tfrac{6}{5} \},&
         &\{ 2, \tfrac{1}{2}, \tfrac{1}{2}, -\tfrac{1}{2}, 1 \},&
         &\{ 2, \tfrac{2}{3}, -\tfrac{2}{3}, \tfrac{10}{3}, -\tfrac{10}{3} \},&
         &\{ 2, \tfrac{2}{3}, \tfrac{4}{3}, \tfrac{4}{3}, -\tfrac{4}{3} \},\\
         &\{ 2, \tfrac{2}{3}, \tfrac{4}{3}, -\tfrac{4}{3}, \tfrac{8}{3} \},&
         &\{ 2, 1, -1, 4, -4 \}. \refstepcounter{equation} \tag{\theequation}
\end{align*}

\section{Analyzing exceptional models in the list}\label{app:b}

The $19$ models in the list with $N_2 > N_0$ and SUSY solutions are exceptions to the Nelson-Seiberg theorem and its generalization.  All of them have at least one pair of fields with opposite R-charges $\pm r$.  But not all such oppositely R-charged field pairs contribute to $N'_\pm$ in Theorem \ref{thm:03}.  One needs to construct the explicit form of the renormalizable superpotential $W$, and check whether they appear only linearly in cubic terms of $W$.  Here we present these exceptions using the notation that $X$ and $Y$ fields have R-charges $2$ and $0$ respectively, $P$ and $Q$ fields have R-charges $r$ and $- r$ with $r > 0, r \ne 2$, and other fields are identified as $A$.

There is only one exceptional model with $N = 4$ fields:
\begin{equation}
\begin{gathered}
\{ R(X), R(A), R(P), R(Q) \} = \{ 2, - 2, 6, - 6 \},\\
W = X (a + b P Q) + \xi X^2 A + \sigma P A^2.
\end{gathered}
\end{equation}
It has one pair of $P$-$Q$ fields which appear only linearly in cubic terms of $W$.  We have $N_2 = N'_\pm = 1$ and $N_Y = 0$.  The SUSY solution is given as
\begin{equation}
X = A
  = 0, \quad
P Q = - \frac{a}{b},
\end{equation}
and the R-symmetry is broken broken by the non-zero VEV's of $P$ and $Q$ fields.  Thus the model is an exception covered by Theorem \ref{thm:03}.

There are $18$ exceptional models with $N = 5$ fields.  $8$ of them have one pair of $P$-$Q$ fields which appear only linearly in cubic terms of $W$:
\begin{gather}
\begin{gathered}
\{ R(X), R(A_1), R(A_2), R(P), R(Q) \} = \{ 2, - 2, - 2, 6, - 6 \},\\
W = X (a + b P Q) + \xi_1 X^2 A_1 + \xi_2 X^2 A_2 + \sigma_1 P A_1^2 + \sigma_2 P A_2^2 + \sigma_{1 2} P A_1 A_2,
\end{gathered}\\
\begin{gathered}
\{ R(X), R(A_1), R(A_2), R(P), R(Q) \} = \{ 2, - 2, \frac{2}{3}, \frac{10}{3}, - \frac{10}{3} \},\\
W = X (a + b P Q) + \xi X^2 A_1 + \lambda A_2^3 + \sigma P A_1 A_2,
\end{gathered}\\
\begin{gathered}
\{ R(X), R(A_1), R(A_2), R(P), R(Q) \} = \{ 2, - 2, 1, 3, - 3 \},\\
W = X (a + b P Q) + \xi X^2 A_1 + \lambda A_2^2 + \sigma P A_1 A_2,
\end{gathered}\\
\begin{gathered}
\{ R(X), R(A_1), R(A_2), R(P), R(Q) \} = \{ 2, - 2, 4, 6, - 6 \},\\
W = X (a + b P Q) + \xi X^2 A_1 + \mu A_1 A_2 + \sigma P A_1^2 + \tau Q A_2^2,
\end{gathered}\\
\begin{gathered}
\{ R(X), R(A_1), R(P), R(Q), R(A_2) \} = \{ 2, - 2, 6, - 6, 10 \},\\
W = X (a + b P Q) + \xi X^2 A_1 + \sigma P A_1^2 + \tau Q A_1 A_2,
\end{gathered}\\
\begin{gathered}
\{ R(X), R(A_1), R(A_2), R(P), R(Q) \} = \{ 2, - 2, 6, 10, - 10 \},\\
W = X (a + b P Q) + \xi X^2 A_1 + \lambda A_1^2 A_2 + \tau Q A_2^2,
\end{gathered}\\
\begin{gathered}
\{ R(X), R(P), R(Q), R(A_1), R(A_2) \} = \{ 2, \frac{2}{9}, - \frac{2}{9}, \frac{2}{3}, \frac{10}{9} \},\\
W = X (a + b P Q) + \lambda A_1^3 + \sigma P A_1 A_2 + \tau Q A_2^2,
\end{gathered}\\
\begin{gathered}
\{ R(X), R(P), R(Q), R(A_1), R(A_2) \} = \{ 2, \frac{1}{3}, - \frac{1}{3}, \frac{2}{3}, 1 \},\\
W = X (a + b P Q) + \lambda A_1^3 + \lambda_2 A_2^2 + \sigma P A_1 A_2.
\end{gathered}
\end{gather}
We have $N_2 = N'_\pm = 1$ and $N_Y = 0$ in these models.  SUSY solutions are given as
\begin{equation}
X = A_1
  = A_2
  = 0, \quad
P Q = - \frac{a}{b},
\end{equation}
and the R-symmetry is broken broken by the non-zero VEV's of $P$ and $Q$ fields.  Thus these models are exceptions covered by Theorem \ref{thm:03}.

$2$ of the $18$ $N = 5$ exceptions have one degenerate pair of $P$-$Q$ with three fields which appear only linearly in cubic terms of $W$:
\begin{gather}
\begin{gathered}
\{ R(X), R(A), R(P_1), R(P_2), R(Q) \} = \{ 2, - 2, 6, 6, - 6 \},\\
W = X (a + b_1 P_1 Q + b_2 P_2 Q) + \xi X^2 A + \sigma_1 P_1 A^2 + \sigma_2 P_2 A^2,
\end{gathered}\\
\begin{gathered}
\{ R(X), R(A), R(P), R(Q_1), R(Q_2) \} = \{ 2, - 2, 6, - 6, - 6 \},\\
W = X (a + b_1 P Q_1 + b_2 P Q_2) + \xi X^2 A + \sigma_1 P A^2.
\end{gathered}
\end{gather}
We have $N_2 = 1$, $N'_\pm = 2$ and $N_Y = 0$ in these models.  SUSY solutions are given as
\begin{equation}
X = A
  = 0, \quad
b_1 P_1 Q + b_2 P_2 Q = 0 \quad
\text{or} \quad
b_1 P Q_1 + b_2 P Q_2 = 0,
\end{equation}
and the R-symmetry is broken broken by the non-zero VEV's of $P$ and $Q$ fields.  Thus these models are exceptions covered by Theorem \ref{thm:03}.

$2$ of the $18$ $N = 5$ exceptions have two pairs of $P$-$Q$ fields:
\begin{gather}
\begin{gathered}
\{ R(X), R(P'), R(Q'), R(P), R(Q) \} = \{ 2, \frac{2}{3}, - \frac{2}{3}, \frac{10}{3}, - \frac{10}{3} \},\\
W = X (a + b P Q + b' P' Q') + \lambda P'^3 + \sigma P Q'^2,
\end{gathered}\\
\begin{gathered}
\{ R(X), R(P'), R(Q'), R(P), R(Q) \} = \{ 2, 1, - 1, 4, -4 \},\\
W = X (a + b P Q + b' P' Q') + \mu P'^2 + \sigma P Q'^2.
\end{gathered}
\end{gather}
In both models, $P$ and $Q$ appear only linearly in cubic terms of $W$, but $P'$ and $Q'$ do not satisfy this condition.  So only $P$ and $Q$ contribute to $N'_\pm$.  We have $N_2 = N'_\pm = 1$ and $N_Y = 0$ in these models.  SUSY solutions are given as
\begin{equation}
X = P'
  = Q'
  = 0, \quad
P Q = - \frac{a}{b},
\end{equation}
and the R-symmetry is broken broken by the non-zero VEV's of $P$ and $Q$ fields.  Thus these models are exceptions covered by Theorem \ref{thm:03}.

$2$ of the $18$ $N = 5$ exceptions have one degenerate pair of $P$-$Q$ with three fields which seem to appear in quadratic terms of $W$:
\begin{gather}
\begin{gathered}
\{ R(X), R(A), R(P_1), R(P_2), R(Q) \} = \{ 2, - 2, 4, 4, -4 \},\\
W = X (a + b_1 P_1 Q + b_2 P_2 Q) + \xi X^2 A + (\mu_1 P_1 + \mu_2 P_2) A,
\end{gathered}\\
\begin{gathered}
\{ R(X), R(A), R(P_1), R(P_2), R(Q) \} = \{ 2, \frac{2}{3}, \frac{4}{3}, \frac{4}{3}, - \frac{4}{3} \},\\
W = X (a + b_1 P_1 Q + b_2 P_2 Q) + (\mu_1 P_1 + \mu_2 P_2) A + \lambda A^3.
\end{gathered}
\end{gather}
A field redefinition
\begin{equation}
P = \frac{\mu_1}{\mu_2} P_1 - P_2, \quad
P' = P_1 + \frac{\mu_2}{\mu_1} P_2
\end{equation}
turns the superpotentials into
\begin{gather}
W = X (a + \frac{b_1 \mu_2 - b_2 \mu_1}{2 \mu_1} P Q + \frac{b_1 \mu_2 + b_2 \mu_1}{2 \mu_2} P' Q) + \xi X^2 A + \mu_1 P' A,\\
W = X (a + \frac{b_1 \mu_2 - b_2 \mu_1}{2 \mu_1} P Q + \frac{b_1 \mu_2 + b_2 \mu_1}{2 \mu_2} P' Q) + \mu_1 P' A + \lambda A^3.
\end{gather}
In both models, $P$ and $Q$ appear only linearly in cubic terms of $W$, but $P'$ does not satisfy this condition.  So only $P$ and $Q$ contribute to $N'_\pm$.  We have $N_2 = N'_\pm = 1$ and $N_Y = 0$ in these models.  SUSY solutions are given as
\begin{equation}
X = A
  = P'
  = 0, \quad
P Q = \frac{- 2 a \mu_1}{b_1 \mu_2 - b_2 \mu_1},
\end{equation}
and the R-symmetry is broken broken by the non-zero VEV's of $P$ and $Q$ fields.  Thus these models are exceptions covered by Theorem \ref{thm:03}.

$2$ of the $18$ $N = 5$ exceptions have one degenerate pair of $P$-$Q$ with three fields which seem to appear quadratically in cubic terms of $W$:
\begin{gather}
\begin{gathered}
\{ R(X), R(P_1), R(P_2), R(Q), R(A) \} = \{ 2, \frac{2}{5}, \frac{2}{5}, - \frac{2}{5}, \frac{6}{5} \},\\
W = X (a + b_1 P_1 Q + b_2 P_2 Q) + (\lambda_1 P_1^2 + \lambda_2 P_2^2 + \lambda_{1 2} P_1 P_2) A + \tau Q A^2,
\end{gathered}\\
\begin{gathered}
\{ R(X), R(P_1), R(P_2), R(Q), R(A) \} = \{ 2, \frac{1}{2}, \frac{1}{2}, - \frac{1}{2}, 1 \},\\
W = X (a + b_1 P_1 Q + b_2 P_2 Q) + \mu A^2 + (\lambda_1 P_1^2 + \lambda_2 P_2^2 + \lambda_{1 2} P_1 P_2) A.
\end{gathered}
\end{gather}
A field redefinition
\begin{equation}
P = \frac{\sqrt{\lambda_{1 2}^2 - 4 \lambda_1 \lambda_2}}{\lambda_{1 2}} P_2, \quad
P' = P_1 + \frac{\lambda_{1 2} - \sqrt{\lambda_{1 2}^2 - 4 \lambda_1 \lambda_2}}{2 \lambda_1} P_2
\end{equation}
turns the superpotentials into
\begin{gather}
W = X (a + b_0 P Q + b_1 P' Q) + \lambda_1 P'^2 A + \lambda_{1 2} P P' A + \tau Q A^2,\\
W = X (a + b_0 P Q + b_1 P' Q) + \mu A^2 + \lambda_1 P'^2 A + \lambda_{1 2} P P' A,\\
\text{where} \quad
b_0 = \frac{\lambda_{1 2} (2 b_2 \lambda_1 - b_1 \lambda_{1 2} + b_1 \sqrt{\lambda_{1 2}^2 - 4 \lambda_1 \lambda_2})}{2 \lambda_1 \sqrt{\lambda_{1 2}^2 - 4 \lambda_1 \lambda_2}}.
\end{gather}
In both models, $P$ and $Q$ appear only linearly in cubic terms of $W$, but $P'$ does not satisfy this condition.  So only $P$ and $Q$ contribute to $N'_\pm$.  We have $N_2 = N'_\pm = 1$ and $N_Y = 0$ in these models.  SUSY solutions are given as
\begin{equation}
X = A
  = P'
  = 0, \quad
P Q = - \frac{a}{b_0}
    = \frac{- 2 a \lambda_1 \sqrt{\lambda_{1 2}^2 - 4 \lambda_1 \lambda_2}}{\lambda_{1 2} (2 b_2 \lambda_1 - b_1 \lambda_{1 2} + b_1 \sqrt{\lambda_{1 2}^2 - 4 \lambda_1 \lambda_2})},
\end{equation}
and the R-symmetry is broken broken by the non-zero VEV's of $P$ and $Q$ fields.  Thus these models are exceptions covered by Theorem \ref{thm:03}.

$2$ of the $18$ $N = 5$ exceptions have one pair of $P$-$Q$ fields which appear in quadratic terms of $W$:
\begin{gather}
\begin{gathered}
\{ R(X), R(A_1), R(P), R(Q), R(A_2) \} = \{ 2, - 2, 4, - 4, 8 \},\\
W = X (a + b P Q) + \xi X^2 A_1 + \sigma P A_1 + \tau Q A_1 A_2,
\end{gathered}\\
\begin{gathered}
\{ R(X), R(A_1), R(P), R(Q), R(A_2) \} = \{ 2, \frac{2}{3}, \frac{4}{3}, - \frac{4}{3}, \frac{8}{3} \},\\
W = X (a + b P Q) + \lambda A_1^3 + \sigma P A_1 + \tau Q A_1 A_2.
\end{gathered}
\end{gather}
We have $N_2 = 1$ and $N'_\pm = N_Y = 0$ in these models, and the condition in Theorem \ref{thm:03} is not satisfied.  But SUSY solutions are given as
\begin{equation}
X = A_1
  = 0, \quad
P Q = - \frac{a}{b}, \quad
A_2 = - \frac{\sigma P}{\tau Q},
\end{equation}
and the R-symmetry is broken broken by the non-zero VEV's of $P$, $Q$ and $A_2$ fields.  Thus these models are exceptions beyond Theorem \ref{thm:03}, but within the scope of~\cite{Nakayama:2023eax}.

\end{document}